\documentclass{aa}

\usepackage{graphicx}
\usepackage[section]{placeins} 
\usepackage{multirow}

\usepackage[varg]{txfonts}  

\usepackage[english]{babel}
\usepackage[babel=true]{csquotes}
\usepackage{amssymb}   

\usepackage{stfloats}  

\bibpunct{(}{)}{;}{a}{}{,}
\begin{document}

	\title{A new sample of X-ray selected narrow emission-line galaxies.}
		\subtitle{I. The nature of optically elusive AGN.}
	
	\author{E. Pons
		\and M. G. Watson}
		
	\institute{Department of Physics \& Astronomy, University of Leicester, Leicester, LE1 7RH, UK\\
		}
		
	\date{Received 24 March 2014 / Accepted 15 July 2014}
	
	\abstract{Using the 3XMM catalogue of serendipitous X-ray sources, and the SDSS-DR9 
	spectroscopic catalogue, we have obtained a new sample of X-ray selected narrow emission line 
	galaxies. The standard optical diagnostic diagram and selection by hard X-ray luminosity expose a 
	mismatch between the optically-based and X-ray-based classifications. The nature of these 
	misclassified \enquote{elusive} AGN can be understood in terms of their broader X-ray and optical 
	properties and leads to a division of this sub-sample into two groups. A little more than half are 
	likely to be narrow-line Seyfert 1s (NLS1s), so misclassified because of the contribution of the 
	Broad Line Region (BLR) to their optical spectra. The remainder have some of the properties of 
	Seyfert 2 (Sy2) AGN; their optical elusiveness can be explained by optical dilution from the host 
	galaxy plus a star-formation contribution and by their underluminous optical emission due to low 
	accretion rates. Because some of the Sy2 sources have very low accretion rates, are unabsorbed, 
	plus the fact that they lack broad optical emission lines, they are good candidates to be 
	\emph{True Sy2} AGN.}
	
	\keywords{Galaxies: active -- Galaxies: Seyfert -- X-rays: galaxies}
	
	\maketitle


\section{Introduction}

   There is strong observational evidence that most, if not all, massive galaxies ($M_\ast > 10^{10} - 
   10^{12} \,\mathrm{M_\odot}$) in the nearby Universe host a central supermassive black hole 
   (SMBH; \citet{Kor95}). Central black holes are less common in low-mass systems, but have still 
   been identified in some low mass and dwarf galaxies \citep{Filippenko03,Barth04,Reines11}. The 
   central sources in galaxies are observed as active galactic nuclei (AGN) as the SMBH grows through a 
   phase of significant mass accretion.

   With large surveys at different wavelengths, various criteria can be used to identify AGN. In the 
   optical, type 2 AGN can be distinguished from star-forming (SF) galaxies on the basis of their 
   emission lines. The most commonly used diagnostic for large sample of galaxies is the BPT diagram 
   introduced by \citet{BPT81}. This considers the emission-line ratios to probe the source of excitation. 
   In the case of photoionization from young and massive O stars (in star-forming galaxies), the low 
   ionization transitions (i.e. the collisionally-excited lines) [NII], [SII], [OIII] and especially [OI] are very 
   weak relative to the Balmer recombination lines. In contrast, for photoionization from an accretion 
   disc around a SMBH, the collisionally-excited lines will be stronger because the photons from an 
   AGN extend to higher energy.

   However, some X-ray selected AGN, selected by their hard X-ray luminosity, will not be classified as 
   AGN by their optical emission line properties. This class of sources has been given a variety of 
   names such as \enquote{elusive} AGN \footnote{In this paper, the term \enquote{elusive} AGN 
   refers to objects with X-ray emission like an AGN but optical emission line ratios like a star-forming 
   galaxy.} \citep{Cac07}, \enquote{optically dull} AGN  \citep{Elvis81,Trump09} or also \enquote{X-ray 
   bright, optically normal galaxies}  (XBONG; \citet{Comastri02}). Such galaxies are particularly interesting 
   because, despite their luminous X-ray emission, they lack clear optical signatures of an AGN. Previous 
   studies have shown that many of these sources are not so special but are indeed normal type 2 AGN 
   diluted by a bright host \citep{Moran02,Cac07}; however about 10-20\% are undiluted 
   \citep{LaFranca02} and so dilution may not be the cause of all optically dull AGN. Another possibility is 
   that the optical emission of elusive AGN has been rendered \enquote{invisible} by obscuration which 
   may be caused be optically-thick gas clouds covering the nuclear source, as suggested by 
   \citet{Comastri02}, or instead by extranuclear gas and dust in the host galaxy \citep{Rigby06}. 
   Alternatively, a fraction of elusive AGN are intrinsically optically weaker than other AGN \citep{Trump09} 
   and can be characterised by unusual properties such as weak emission from the accretion disk (e.g. 
   radiatively inefficient accretion flow RIAF, \citet{Yuan04}). Finally more recent work by  \citet{Castello12} 
   proposes that the optical misclassification of X-ray bright AGN through emission lines may be due to the 
   contamination of the narrow-line galaxy samples by narrow-line Seyfert 1 (NLS1), a subclass of 
   unobscured AGN which have narrower line width (usually between 500 and 2000 $\mathrm{km.s^{-1}}$) 
   and a $\mathrm{[OIII]/H}\beta$ line ratio smaller than 3 \citep{Boller96,Botte04}.
   
   In this paper we investigate a new sample of Seyfert 2 (Sy2) galaxies with conflicting X-ray and 
   optical classifications. After identifying possible NLS1, we consider three explanations for optical 
   dullness of the remaining Sy2 galaxies in the sample: (1) obscuration by an dusty torus (Compton-
   thick AGN, $N_H > 10^{24} \, \mathrm{cm^{-2}}$); (2) optical dilution from the host galaxy star-light; 
   (3) weak emission from the accretion disc related low accretion rates.
   
   Some of the AGN in our sample may belong to a more exotic class of AGN which are inconsistent 
   with the predictions of the Unified Model \citep{Antonucci93, Urry95}. According to this scheme, AGN 
   are classified into two classes based on the presence (type 1) or absence (type 2) of broad emission 
   lines in their optical spectra. This means that the central black hole (BH) is viewed directly (type 1) or 
   is obscured by a dusty torus (type 2). However, type 2 AGN without an hidden broad line region (BLR) 
   have been found \citep{Bian12}; these are not predicted by the model. These sources, often called 
   \emph{True Seyfert 2}, may be observed while accreting at low Eddington rates, in agreement with 
   models which predict that the BLR disappears below a certain accretion rate threshold \citep{Yuan07} 
   and the absence of observed broad-line AGN below a limiting accretion rate (corresponding to an 
   Eddington ratio smaller than $10^{-2}$) \citep{Trump11}.
 
   The aim of this paper is to understand the nature of a new sample of AGN which have conflicting 
   properties in the optical and X-ray band. This sample also allows us to find peculiar AGN with 
   unexpected properties.
   The paper is structured as follows: section \ref{sec:sample} describes the source selection; the X-
   ray and optical properties are detailed in section \ref{sec:pties}; and the section 
   \ref{sec:misclassified} discusses the most likely explanations for optical dullness.

\section{Narrow emission-line galaxies sample}
\label{sec:sample}
	\subsection{Source selection}
	The 3XMM catalogue of serendipitous X-ray sources (Watson et al. 2014; in preparation) is 
	currently the largest X-ray catalogue containing over 370,000 discrete entries drawn from over 
	7000 pointed XMM-Newton observations covering around 2\% of the sky. About 45\% of  3XMM 
	catalogue sources lie in the region covered by the Sloan Digitial Sky Survey (SDSS). To create 
	our sample of X-ray selected galaxies we carried out a positional cross-match of the 3XMM 
	catalogue with SDSS-DR9, the ninth SDSS data release, focusing exclusively on galaxies 
	with SDSS spectroscopy ($\sim 1,690,000$ objects). This allows us to create a large sample of 
	sources with useful X-ray and optical data. 
		
	From SDSS we selected galaxies with a spectroscopic classification of \enquote{GALAXY} or 
	\enquote{QSO} and which are \enquote{science primary} objects (i.e. those that have the best 
	available spectra). The cross-matched X-ray/optical sample is obtained by cross-correlating the 
	SDSS spectroscopic galaxies sub-sample with the 3XMM catalogue. SDSS galaxies are 
	considered as \textbf{potential} counterparts to the X-ray sources if the separation between the 
	optical and X-ray positions is smaller than 10'' and for which the normalised separation (i.e. the 
	ratio of the separation to the X-ray position error) is smaller than 4 \citep{Pineau11}. 
	Because we restrict our sample to SDSS spectroscopic objects, which are significantly brighter 
	and thus have lower sky density than  galaxies in the SDSS photometric catalogue, spurious 
	matches become negligible even with these generous limits. In addition, X-ray sources are 
	required to be point-like (3XMM catalogue parameter SC\_EXTENT < 5; thus removing galaxy 
	clusters) and to have a relatively high detection significance (catalogue parameter SC\_DET\_ML 
	> 15). This results, at this stage, in a cross-matched catalogue of about 4900 X-ray objects with 
	SDSS spectroscopic data.
	
	From this sample, narrow emission-line galaxies are selected by requiring the full-width at half 
	maximum (FWHM) of the Balmer lines to be smaller than 1000 $\mathrm{km.s^{-1}}$ \citep[e.g.][]
	{Cac08}; this threshold rejects all \emph{bona fide} broad-line objects and reduces the 
	contamination by conventional Narrow Line Seyfert 1 (NLS1). Narrow emission-line galaxies 
	represent 31\% of the whole cross-matched catalogue, corresponding to 1555 sources.
	
	Additional derived products for galaxies in the SDSS spectroscopic database are available from 
	the MPA-JHU group. These are the GALPSPEC measurements (described in \citet{Bri04, 
	Kauff03a, Trem04}). The derived spectral products include emission line fluxes corrected for 
	Galactic reddening (using the maps of \citet{Schlegel98}). In addition, through fitting of the 
	observed spectrum with galaxy spectra models, the derived products provide estimates of 
	emission line strengths after subtraction of stellar absorption line components. This is important 
	because the SDSS aperture of the spectroscopic fibre is 3" and thus the spectra include not only  the 
	nucleus but have also host galaxy contribution. Moreover, using stellar population models, the 
	derived products include a variety of galaxy parameters (e.g. stellar masses, stellar and gas 
	kinematics, velocity dispersions \ldots). We choose to keep only sources for which these improved 
	derived spectral products were available. Furthermore we removed from our sample those 
	galaxies which are, SDSS \enquote{X-ray targets} (those selected from ROSAT, Chandra and 
	XMM catalogues), together with objects from the SDSS BOSS sample (most of which, in our 
	sample, are X-ray selected or are selected because they other particular properties). Moreover 
	sources which are the targets of XMM-Newton observations are also removed. We have made this 
	selection because these SDSS and XMM \enquote{targets} would represent a biased subsample. 
	\footnote{An unbiased sample is required for future analysis using the same dataset.}
	After the new selections have been applied the final sample consists of 1514 objects, 
	corresponding to 97\% of the narrow emission-line galaxies.

	Finally, to have a trustworthy sample classification, galaxies are only considered if they have a 
	reasonable quality optical spectrum (signal to noise ratio of the whole spectrum $S/N > 3$) and a 
	redshift $z<0.4$ (so that all of the four lines H$\beta$, [OIII]$\lambda$5007, H$\alpha$ and [NII]$
	\lambda$6584 are covered). We further require that these lines are in emission with a reliable 
	measurement; specifically the line ratios $\mathrm{[OIII]/H}\beta$ and $\mathrm{[NII]/H}\alpha$ 
	must be three times larger than the error on the ratio.
	
	Using these criteria, the final sample of X-ray selected narrow-emission line galaxies (NELGs) 
	includes 797 X-ray sources with good SDSS spectroscopy and a median redshift of 0.15.

	\subsection{Classification}
		\subsubsection{Optical classification}	
		
	The BPT diagnostic diagram is used to classify the NELGs sample on the basis of their optical 
	emission properties. The classification scheme that best distinguishes between AGN and star-
	forming galaxies uses the $\mathrm{[OIII]/H}\beta$ vs. $\mathrm{[NII]/H}\alpha$ emission-line 
	ratios \citep{Stasi06}. The other main virtues of this scheme are that the lines used are relatively 
	strong, lie in an easily accessible region of the optical spectrum and the line ratios are relatively 
	insensitive to reddening because of their close separation in wavelength.
	
	\begin{figure*}
		\centering
		\begin{tabular}{cc}
		   \includegraphics[width=8.5cm]{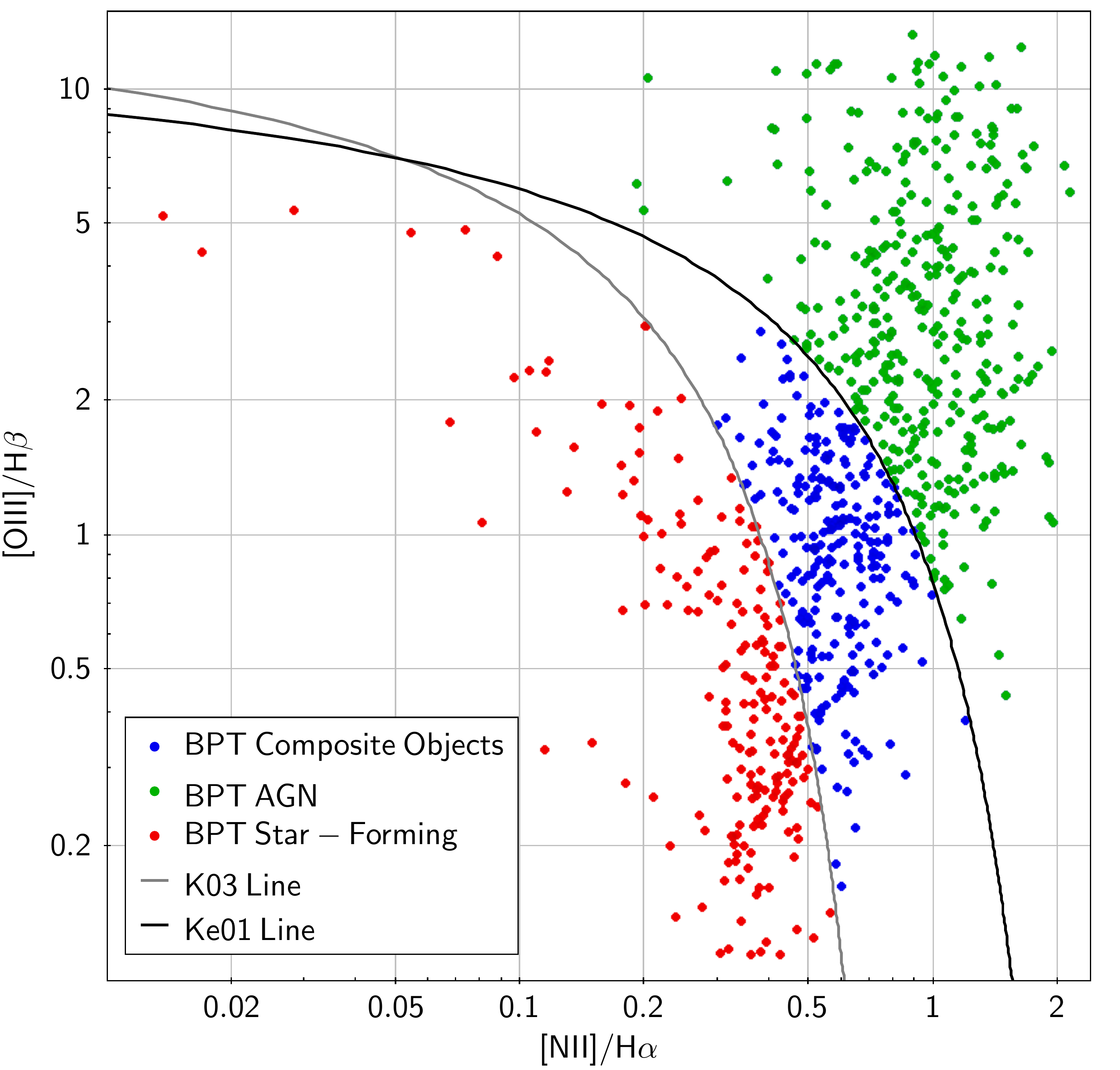}&
	           \includegraphics[width=8.5cm]{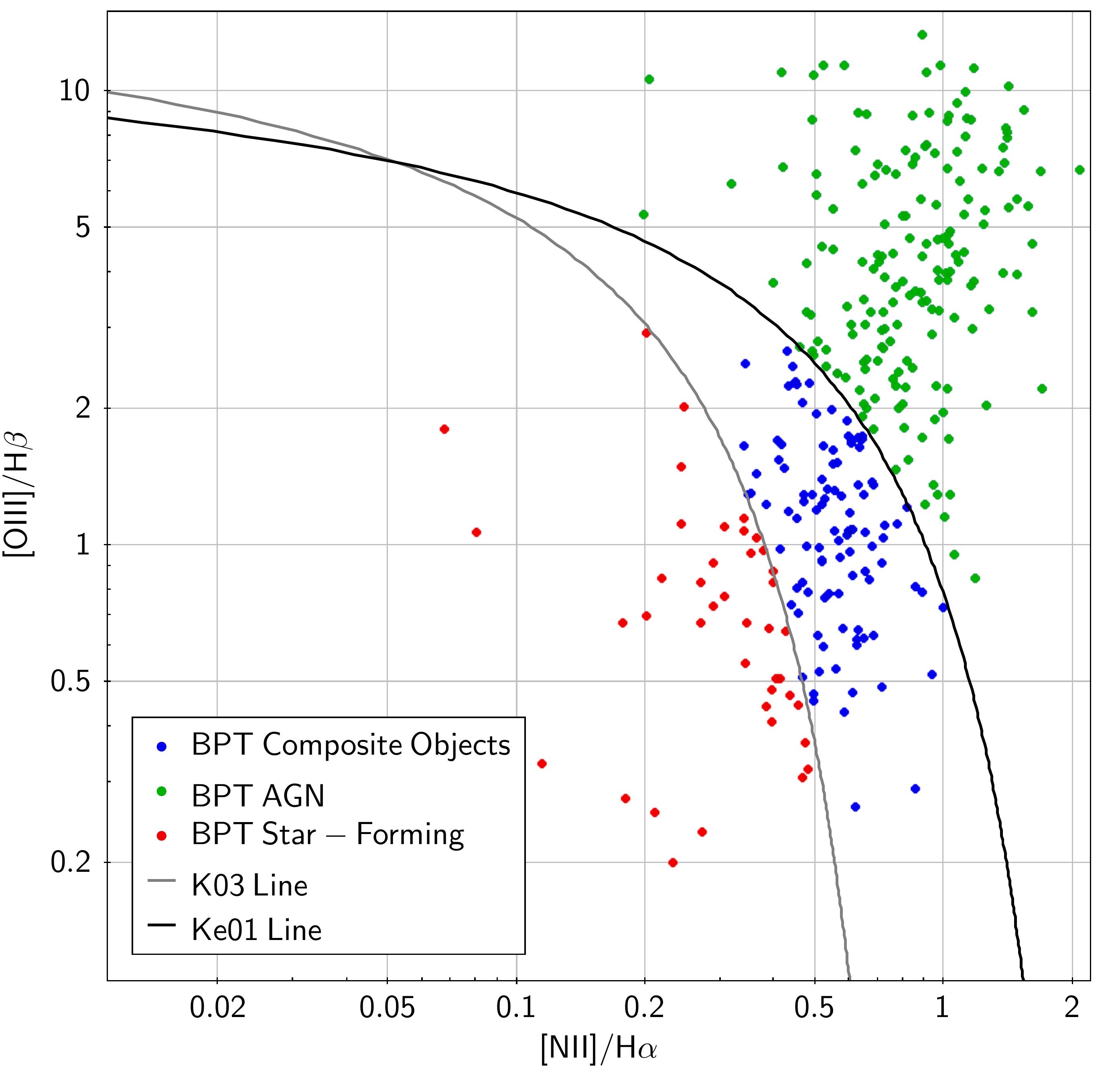}\\
	        \end{tabular}
	   \caption{BPT emission line diagnostic diagrams. The K03 line in grey separates the 
	   SF galaxies and composite object regions, while the Ke01 demarcation line in black distinguishes 
	   between optically-classified AGN and composite objects. \textbf{Left panel}: NELG sample 
	   which consists of 797 sources (45\% of BPT-AGN (green symbols), 31\% composite objects 
	   (blue symbols) and 24\% in the SF region (red symbols). \textbf{Right panel}: NELG sample with 
	   high hard X-ray luminosity ($L_{HX}>10^{42} \,\mathrm{erg.s^{-1}}$). There are 172 sources (54\% 
	   of the sample) classified as BPT-AGN (green symbols), 103 sources (blue symbols) 
	   classified as composite objects and 41 sources (13\%) in the SF region of the diagram (red 
	   symbols; our elusive AGN).}
	   \label{BPTdiag}
	\end{figure*}

	Two demarcation lines are used here for the classification. The first is the theoretical division 
	from \citet[hereafter Ke01]{Kewley01}, which uses a combination of photoionization and stellar 
	population synthesis models to place a theoretical conservative lower limit on the true number of 
	AGN:
	$$ \log{\left( \frac{[OIII]\lambda5007}{H\beta}\right)} =\frac{0.61}{\log{\left(\frac{[NII]\lambda6584}
	{H\alpha}\right)}-0.47}+1.19 $$
	We also use an empirical line ratio based on the data from \citet[hereafter K03]{Kauff03b} that 
	places a conservative limit on the true number of SF galaxies: 
	$$ \log{\left( \frac{[OIII]\lambda5007}{H\beta}\right)} =\frac{0.61}{\log{\left(\frac{[NII]\lambda6584}
	{H\alpha}\right)}-0.05}+1.3 $$
	The objects that lie below the K03 line are optically classified as SF galaxies (BPT-SF), the ones that 
	are above the Ke01 line are classified as AGN (BPT-AGN). The objects that lie between these two 
	lines are called Composite objects because their optical spectra may include a contributions from 
	both star-formation and an active nucleus. The classification for the whole NELG sample is shown in 
	the left panel of the Fig. \ref{BPTdiag}.
	
		\subsubsection{X-ray classification}
		
	An alternative way of classifying galaxies as AGN uses an empirical X-ray luminosity threshold. 
	The rest-frame hard X-ray luminosity in the 2 - 10 keV band ($\mathrm{L_{HX}}$) is computed 
	from the measured flux, making the simple assumption that the X-ray spectrum has a power-law 
	form with a photon index $\mathrm{\Gamma}$ of 1.7. The luminosity estimated this way is 
	relatively insensitive to changes in the assumed photon index.  Choosing sources with $L_{HX}
	>10^{42} \,\mathrm{erg.s^{-1}}$ is known to remove galaxies with X-ray emission powered by 
	mechanisms other than nuclear activity \citep{Fabbiano89, Mushotzky04}. This empirical limit 
	works because for a star-forming galaxy to have such a high X-ray luminosity requires a star 
	formation rate (SFR) of at least $200 \,\mathrm{M_\odot.yr^{-1}}$ (e.g. relation from \citet{Ran03}; 
	$L_{HX}=SFR[\mathrm{M_\odot \cdot yr^{-1}}]\cdot 5 \cdot 10^{39} \,\mathrm{erg.s^{-1}}$). Such a 
	high SFR corresponds to that of powerful Ultra Luminous Infrared Galaxies (ULIRGs; 
	$SRF_{min,ULIRG} \sim 100 \,\mathrm{M_\odot.yr^{-1}}$), objects which have a low space density, 
	and thus are unlikely to be a significant contaminant at z<0.4. So selection with a threshold $L_{HX} 
	> 10^{42} \,\mathrm{erg.s^{-1}}$ should produce a sample that only contains AGN. However, among 
	the 316 NELGs with X-ray luminosity above this threshold, there remain 41 sources (13\%) that are 
	optically classified as SF galaxies. We refer to these as elusive AGN (see Fig.\ref{BPTdiag}, right 
	panel).	
	
	The SFR provide in the SDSS GALSPEC catalogue ($\mathrm{SFR_{SDSS}}$)  		
\section{X-ray vs. optical properties}
\label{sec:pties}

	\begin{figure}
	\centering
	   \includegraphics[width=\hsize]{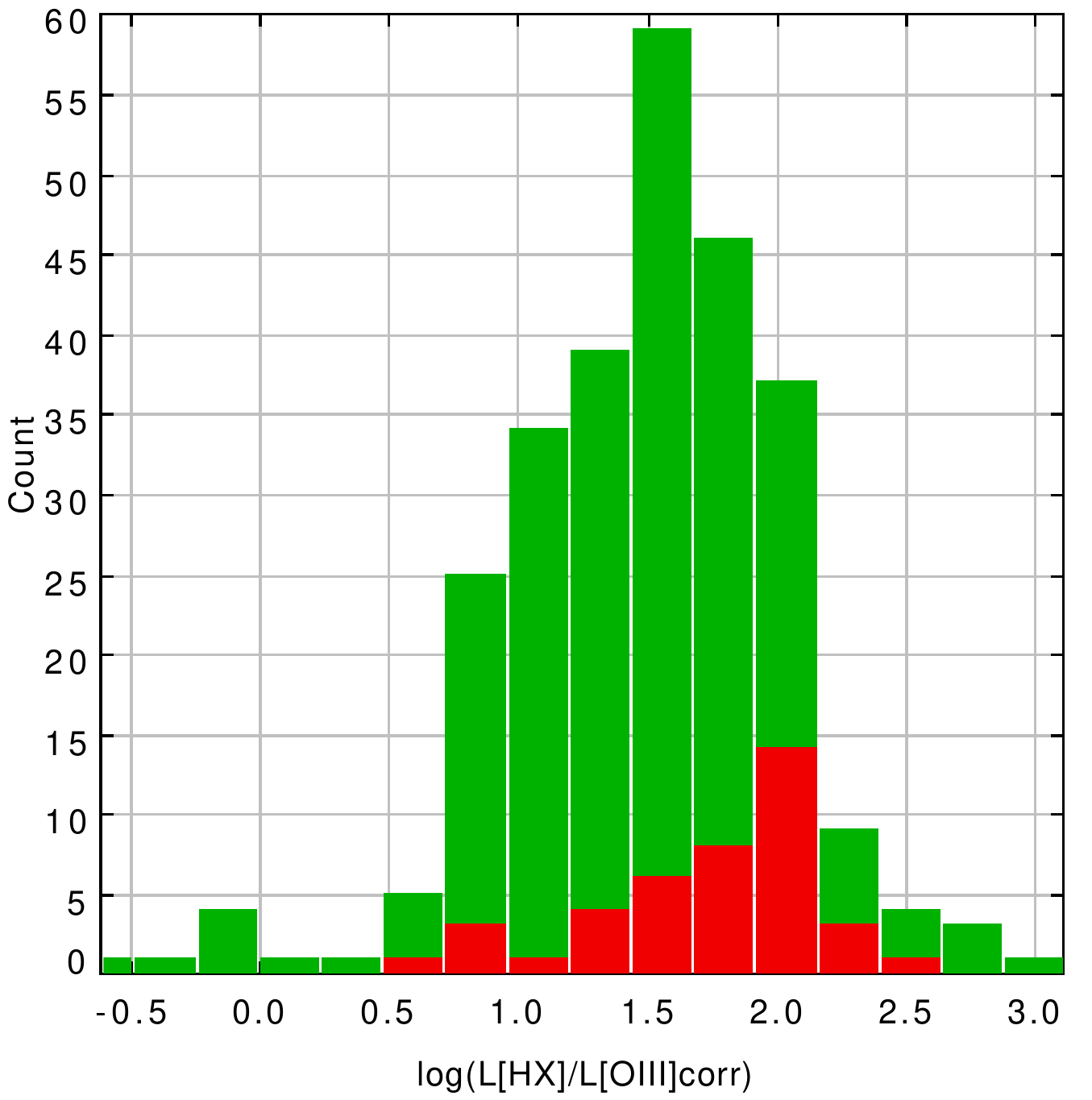}
	   \caption{$\mathrm{L_{HX}/L_{[OIII],corr}}$ distribution for the BPT-AGN and composite objects 
	   (green) and the elusive AGN (red). The latter sources have a higher mean ratio 
	   compared to the other AGN in the sample.}
	   \label{L_dist}
	\end{figure}

   In this section we discuss the X-ray and optical properties of the elusive AGN.
	\subsection{X-ray vs. [OIII] luminosity}
	\label{XvsOpPties}
	
	We compare the hard X-ray and [OIII] luminosities using [OIII] luminosities which are corrected for 
	intrinsic absorption using the Balmer decrement, assuming an intrinsic ratio $\left(H\alpha / H\beta 
	\right)_0$ of 3.1 in the NLR \citep[see][]{OstFer06} and a $\mathrm{H\beta / H\alpha}$ colour 
	index for extinction $\mathrm{\alpha}=2.94$ \citep{Bassani99}:
	$$f_{corr}=f_{obs} \cdot \left(\frac{H\alpha/H\beta}{\left(H\alpha / H\beta \right)_0}\right)^\alpha$$
	Compared to the BPT-AGN and composite objects in the sample, the elusive AGN 
	have a mean thickness parameter T ($\mathrm{T = L_{HX}/L_{[OIII]}}$) which is larger 
	($\mathrm{\log_{10} T \sim 2.1}$) than the other sources ($\mathrm{\log_{10} T \sim 1.5}$); but 
	still in the typical range of values for Seyfert galaxies ($\mathrm{-1 \lesssim \log_{10} T \lesssim 
	2}$; \citet{Bassani99}) (see Fig.\ref{L_dist}). This is consistent with the idea that some of the 
	AGN in our sample could have anomalously low [OIII] flux \citep[see also][]{Trouille10}.

	\subsection{Emission line width vs. X-ray spectral information}
	A crude indicator of the X-ray spectral properties can be deduced from the Hardness Ratio (HR) 
	diagram (also called X-ray colour-colour diagram) which simply compares count rates in 
	adjacent energy bands. The HR is defined in terms of the observed counts rates $R$ in energy  
	bands $n$ and $n+1$ as follows:
	$$HRn=\left(R_{n+1}-R_n\right)/\left(R_{n+1}+R_n\right) \quad\mbox{for}\quad n=1 - 4$$
	where the five energy bands are $0.2-0.5$ keV, $0.5-1.0$ keV, $1.0-2.0$ keV, $2.0-4.5$ keV and 
	$4.5-12$ keV.
	
	\begin{figure}
	\centering
	   \includegraphics[width=\hsize]{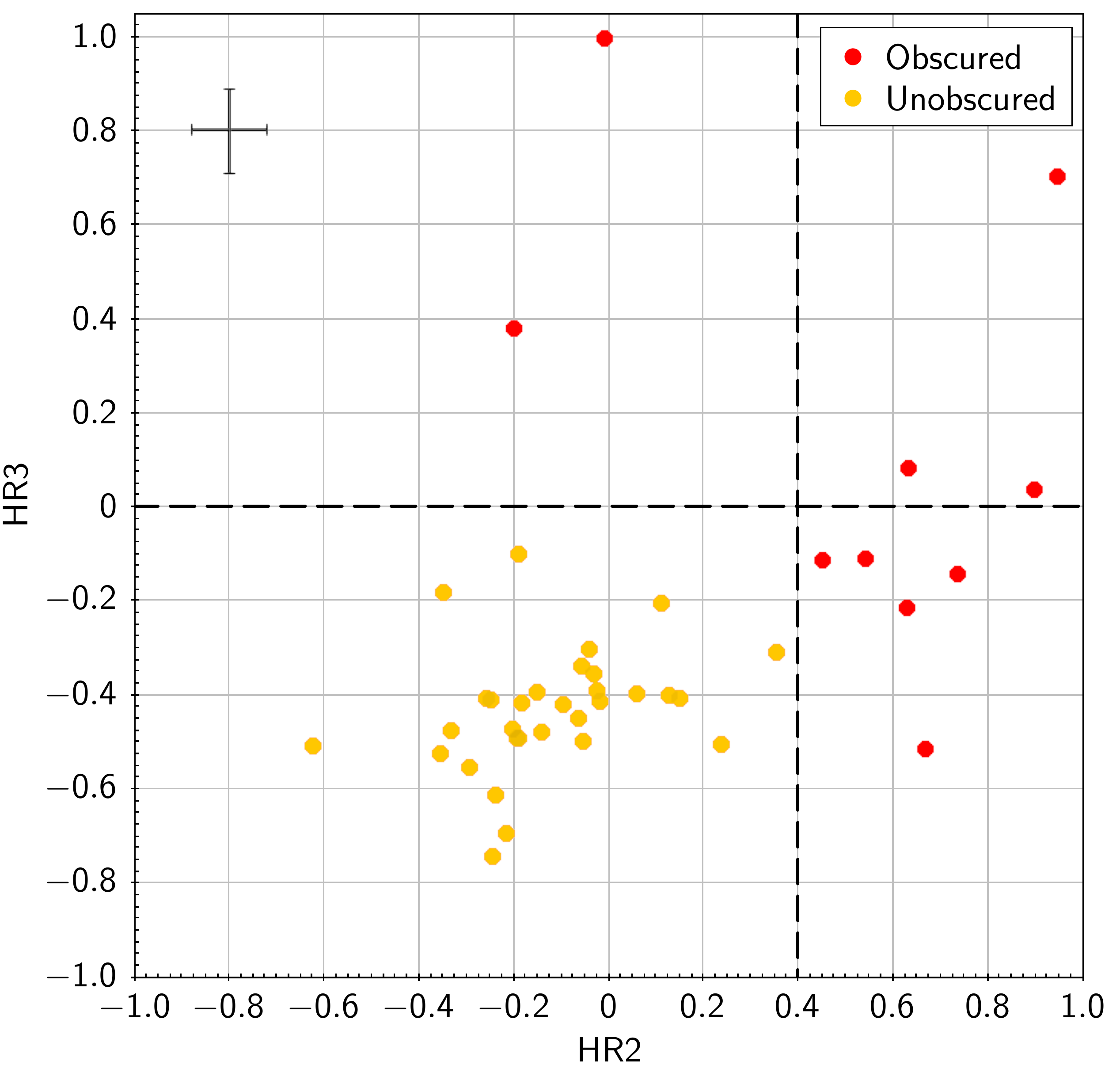}
	   \caption{X-ray colour-colour diagram (or HR diagram) of the elusive AGN. Gives a crude indication 
	   of the intrinsic X-ray spectral shape and level of absorption. The majority 
	   of the sources have a low HR (in yellow) consistent with being unabsorbed or having low 
	   absorption column density $N_H \lesssim 10^{22}\,\mathrm{cm^{-2}}$. The errors for both HR2 
	   and HR3 range from about 0.01 to 0.2, with a mean values of respectively 0.08 and 0.09. The 
	   size of the typical (mean) errors is indicated by the black error-bars shown in the top left of the 
	   diagram.}
	   \label{HRdiag}
	\end{figure}

	The HR3 vs. HR2 diagram, shown in Fig.\ref{HRdiag}, gives an approximate indicator of the 
	intrinsic X-ray spectral shape and is also sensitive to the level of absorption. AGN with intrinsic 
	absorption have a X-ray spectrum much harder than an unabsorbed one because the soft X-ray 
	emission is differentially attenuated by the absorber. The dominant source of variation in the HR 
	values is absorption \citep[see][Fig. 2]{Wang04}. So an unabsorbed X-ray spectrum has HR2 and 
	HR3 values lower than an absorbed one for a wide range of assumed continuum shapes. A typical 
	type 1 AGN has a  HR2 value $\sim0$ and HR3$<0$;  an increase in absorption above $10^{22}\,
	\mathrm{cm^{-2}}$ would increase the HR values to HR2$\sim0.4$ and HR3$\sim0$ \citep[see]
	[Fig.12]{Watson09}. About 73\% of the elusive AGN have low hardness ratios ($
	\mathrm{HR2 < 0.4}$ and $\mathrm{HR3 < 0}$), consistent with being unabsorbed or having low 
	absorption.
	
	The selected narrow-line AGN, even with a threshold of $\mathrm{1000\,km^{-1}}$ for the 
	FWHM of the Balmer lines, could still be contaminated by NLS1. Indeed it has previously been 
	noted that NLS1 can lie in the star-forming region of the BPT diagram \citep{Osterbrock85, 
	Rodriguez00} and that the lower end of their FWHM distribution can extend down to $
	\mathrm{500-600\,km\,s^{-1}}$ \citep{Boller96}. 
	
\begin{figure}
	   \centering
	   \includegraphics[width=\hsize]{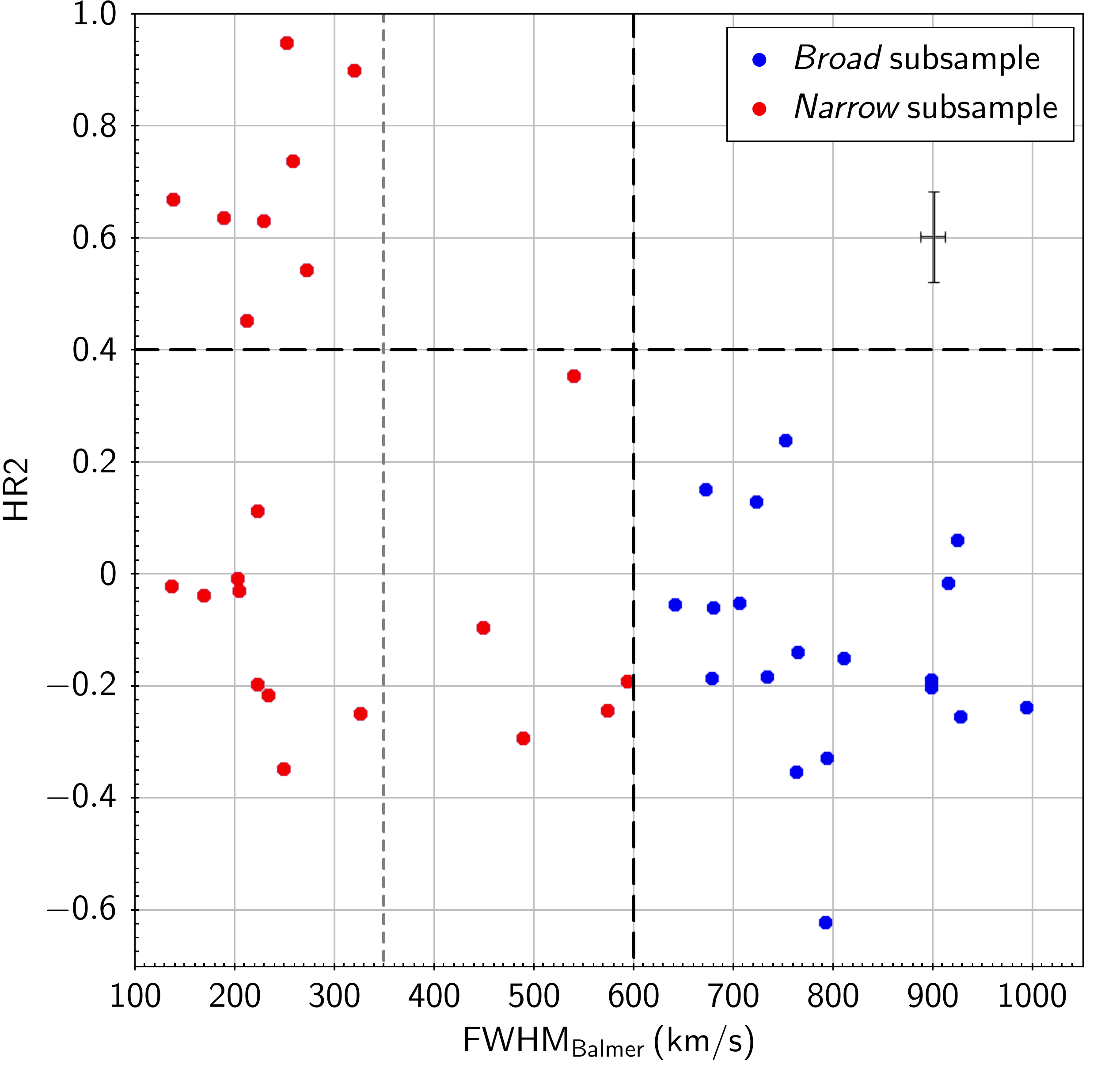}
	   \caption{HR2 vs. Balmer velocity width of elusive AGN. A total of 19 sources (blue 
	   symbols) do not show clear signs of absorption and have broader line widths; they are NLS1 
	   candidates (\emph{broad} sub-sample). The remaining 22 sources with narrower line widths 
	   (red symbols; \emph{narrow} sub-sample) show different properties, only 41\% of them 
	   exhibit clear signs of absorption. The size of the typical (mean) errors is indicated by the black 
	   error-bars shown in the top right of the diagram.}
	   \label{FWHMvsHR}
	\end{figure}

	If we divide our elusive sample into \enquote{broad} line and \enquote{narrow} line 
	sub-samples (hereafter referred as \emph{narrow} and \emph{broad} sub-samples 
	respectively) by taking a $\mathrm{FWHM_{Balmer}}$ limit of $\mathrm{600\,km.s^{-1}}$, the 19 
	broader line sources ($\mathrm{FWHM_{Balmer }> 600\,km.s^{-1}}$), all have a low HR (indicating 
	low absorption) and so are candidates to be NLS1. On the other hand the nature of the sources with 
	lower velocity widths is at this point unclear and only 41\% of them show clear signs of absorption 
	(i.e. have high HR; the others \enquote{narrow} sources with low HR are referred as the 
	\emph{unobscured narrow} sub-sample). Most of them also have a velocity width smaller than 350 
	$\mathrm{km\,s^{-1}}$ corresponding to characteristic values of host galaxy lines and so do not seem 
	to be dominated by an AGN in the optical. Five sources have velocity width broader than expected for 
	a host galaxy, so may be AGN dominated, but still below the broad limit chosen to distinguish 
	between NLS1 and Sy2 ($\mathrm{FWHM_{balmer}\sim 400-600 \,km.s^{-1}}$) (see Fig.
	\ref{FWHMvsHR}). These five sources have an ambiguous classification based on their 
	Balmer velocity line width only; they can be classical AGN dominated X-ray Sy2 or NLS1 but with 
	very low line widths compared to common sources of this class. The separation between the 
	\emph{broad} and \emph{narrow} sub-samples for these sources require further analysis and will be 
	discussed in section \ref{sec:ForbPerm}. At this stage possible NLS1 in the elusive sample 
	are assumed to have line FWHM greater than 600 $\mathrm{km.s^{-1}}$and so belong to the 
	\emph{broad} sub-sample.
	
	\begin{table*}
		\centering
		\caption{Observational parameters and classification of the elusive sources}
		\label{TabParam}

\begin{tabular}{r c c r c c l c c c}
	\hline \hline
	3XMM &IAU Name &\multirow{2}*{OBSID} &$t_{exp}$ &EPIC-pn &$N_{H,gal}$ &Sub- &\multirow{2}*{z\tablefootmark{(2)}} &$L_{HX}$ &$\mathrm{M_{r}}$\tablefootmark{(3)}\\
	 SRCID &(3XMM $\cdots$) & &(ks) &counts &($\cdot 10^{20}\,\mathrm{cm^{-2}}$) &sample\tablefootmark{(1)} & &($\cdot 10^{42}\,\mathrm{erg.s^{-1}}$) &(mag)\\ \hline 
	 1780       &J121048.0+393745   &0112830201   &58.9   &$331\pm14$\tablefootmark{$\diamond$}   &2.13   &$\mathcal{N}1$ $(n\mathcal{N}1)$  &0.20   &$6.17\pm1.21$   &-22.3\\
	 3161       &J094540.4+094749   &0150970301   &23.4   &$154\pm15$\tablefootmark{$\star$}           &2.50   &$\mathcal{N}$ $(n\mathcal{N})$      &0.16   &$6.44\pm1.56$   &-22.2\\
	 3173       &J103342.7+392926   &0506440101   &87.8   &$498\pm26$\tablefootmark{$\diamond$}   &1.31   &$\mathcal{N}$ $(n\mathcal{N})$      &0.07   &$1.99\pm0.31$   &-21.5\\
	 7207       &J095921.2+024030   &0302351601   &30.5   &$2017\pm56$                                             &1.76   &$\mathcal{B}$ $(n\mathcal{B})$      &0.26   &$16.97\pm1.45$ &-23.0\\
	 18567     &J001804.8-005141    &0403760701   &27.1   &$1271\pm39$                                             &2.99   &$\mathcal{B}$ $(n\mathcal{B})$      &0.19   &$7.27\pm1.29$   &-19.5\\
	 18747     &J135810.3+653300   &0305920601   &12.0   &$463\pm26$                                               &1.38   &$\mathcal{B}$ $(n\mathcal{B})$      &0.23   &$8.84\pm1.65$   &-21.8\\
	 \tablefootmark{$\ast$}28357     &J134834.9+263109   &0097820101   &34.3   &$42599\pm218$                                         &1.21   &$\mathcal{B}$ $(n\mathcal{B})$      &0.06   &$8.24\pm0.16$   &-20.9\\
	 28899     &J094240.9+480017   &0201470101   &11.9   &$393\pm22$                                               &1.19   &$\mathcal{N}1$ $(n\mathcal{B})$    &0.20   &$2.61\pm1.22$   &-21.1\\
	 29911     &J100234.8+024253   &0203360101   &26.0   &$416\pm25$                                               &1.78   &$\mathcal{B}$ $(n\mathcal{B})$      &0.20   &$2.79\pm0.48$   &-21.4\\
	 36405     &J231858.2+000621   &0305600601   &16.0   &$169\pm15$\tablefootmark{$\star$}           &3.90   &$\mathcal{B}$ $(n\mathcal{B})$      &0.32   &$12.84\pm7.03$ &-21.8\\
	 47793     &J112758.2+583558   &0112810101   &20.6   &$123\pm15$\tablefootmark{$\star$}           &0.91   &$\mathcal{N}1$ $(n\mathcal{N}1)$  &0.18   &$1.32\pm0.53$   &-21.9\\
	 88476     &J080455.5+241124   &0203280201   &6.3     &$311\pm19$                                               &3.92   &$\mathcal{N}1$ $(n\mathcal{B})$    &0.14   &$4.53\pm1.22$   &-22.0\\
	 93560     &J015727.2-004041    &0303110101   &3.5     &$255\pm20$\tablefootmark{$\star$}           &2.61   &$\mathcal{B}$ $(n\mathcal{B})$      &0.39   &$47.50\pm21.50$ &-22.3\\
	 93640     &J015804.7-005222    &0303110101   &3.4     &$356\pm22$\tablefootmark{$\star$}           &2.50   &$\mathcal{N}1$ $(n\mathcal{N}1)$  &0.08   &$1.18\pm0.50$   &-19.7\\
	 115294   &J004335.9+001455   &0303562201   &4.4     &$155\pm14$\tablefootmark{$\star$}           &1.72   &$\mathcal{B}$ $(n\mathcal{B})$      &0.11   &$5.32\pm2.73$   &-21.4\\
	 120992   &J002253.2+001659   &0407030101   &27.0   &$207\pm20$\tablefootmark{$\star$}           &2.75   &$\mathcal{N}$ $(n\mathcal{N})$      &0.21   &$13.31\pm2.42$ &-21.4\\
	 126294   &J011929.0-000839    &0605391101   &7.3     &$596\pm26$\tablefootmark{$\diamond$}   &3.35   &$\mathcal{B}$ $(n\mathcal{B})$      &0.09   &$8.95\pm2.30$   &-20.8\\
	 155684   &J030634.0-001355    &0201120101   &12.1   &$145\pm19$\tablefootmark{$\star$}           &6.14   &$\mathcal{N}$ $(n\mathcal{N})$      &0.22   &$1.71\pm0.71$   &-22.2\\
	 155872   &J030705.8-000009    &0142610101   &40.3   &$1355\pm42$                                             &6.27   &$\mathcal{B}$ $(n\mathcal{B})$      &0.27   &$10.20\pm1.62$ &-21.0\\
	 266064   &J212929.6+000102   &0093030201   &34.7   &$178\pm20$\tablefootmark{$\star$}           &3.67   &$\mathcal{N}$ $(n\mathcal{N})$      &0.13   &$1.95\pm0.27$   &-21.4\\
	 266125   &J213026.2-000204    &0093030201   &34.6   &$144\pm22$\tablefootmark{$\star$}           &3.65   &$\mathcal{N}1$ $(n\mathcal{N}1)$  &0.14   &$1.77\pm0.89$   &-21.7\\
	 275370   &J104549.6+213106   &0128531601   &76.5   &$62\pm13$\tablefootmark{$\star$}             &1.83   &$\mathcal{N}1$ $(n\mathcal{N}1)$  &0.14   &$1.17\pm0.62$   &-21.6\\
	 277671   &J110738.9+520644   &0304071201   &5.9     &$167\pm16$\tablefootmark{$\star$}           &0.94   &$\mathcal{N}$ $(n\mathcal{N})$      &0.18   &$17.49\pm4.78$ &-22.1\\
	 282027   &J102812.6+293223   &0301650401   &7.8     &$115\pm13$\tablefootmark{$\star$}           &1.91   &$\mathcal{B}$ $(n\mathcal{B})$      &0.29   &$16.01\pm12.10$ &-21.7\\
	 291676   &J115648.6+064753   &0301651801   &4.0     &$305\pm19$\tablefootmark{$\star$}           &1.14   &$\mathcal{N}1$ $(n\mathcal{B})$    &0.15   &$3.18\pm2.85$   &-22.2\\
	 294033   &J111443.6+525834   &0143650901   &3.7     &$885\pm31$                                               &0.97   &$\mathcal{B}$ $(n\mathcal{B})$      &0.08   &$3.60\pm0.86$   &-21.4\\
	 297574   &J112405.1+061248   &0103863201   &5.1     &$896\pm33$                                               &4.61   &$\mathcal{B}$ $(n\mathcal{B})$      &0.27   &$28.25\pm4.80$ &-22.5\\
	 303293   &J091440.9+435716   &0604680201   &14.3   &$148\pm18$\tablefootmark{$\star$}          &1.45   &$\mathcal{N}1$ $(n\mathcal{N}1)$   &0.19   &$2.74\pm1.36$   &-22.3\\
	 305003   &J084058.1+383300   &0502060201   &16.1   &$991\pm34$                                               &3.15   &$\mathcal{N}1$ $(n\mathcal{N}1)$  &0.12   &$3.19\pm0.61$   &-21.5\\
	 321587   &J094057.1+032400   &0306050201   &22.3   &$5213\pm76$                                             &2.99   &$\mathcal{B}$ $(n\mathcal{B})$      &0.06   &$4.91\pm0.32$   &-20.8\\
	 324308   &J134235.6+262533   &0108460101   &25.6   &$1002\pm33$\tablefootmark{$\diamond$} &1.03   &$\mathcal{B}$ $(n\mathcal{B})$      &0.06   &$3.78\pm0.44$   &-20.9\\
	 337999   &J151128.9+561317   &0670650401   &7.4     &$655\pm29$\tablefootmark{$\star$}           &1.38   &$\mathcal{N}1$ $(n\mathcal{B})$    &0.15   &$16.48\pm2.91$ &-22.4\\
	 339379   &J151525.0+042146   &0653810601   &7.6     &$263\pm22$\tablefootmark{$\star$}           &3.57   &$\mathcal{N}1$ $(n\mathcal{N}1)$  &0.10   &$3.91\pm0.54$   &-21.5\\
	 341733   &J141449.5+361240   &0148620101   &13.9   &$1217\pm17$                                             &1.01   &$\mathcal{B}$ $(n\mathcal{B})$      &0.18   &$10.59\pm1.19$ &-21.4\\
	 348033   &J123155.5+200333   &0301450201   &24.1   &$67\pm11$\tablefootmark{$\star$}             &2.58   &$\mathcal{N}$ $(n\mathcal{N})$      &0.11   &$1.51\pm0.74$   &-20.8\\
	 350309   &J123748.5+092323   &0504100601   &17.6   &$2224\pm51$                                             &1.47   &$\mathcal{B}$ $(n\mathcal{B})$      &0.12   &$11.24\pm2.14$ &-21.7\\
	 350410   &J123719.3+114915   &0112840101   &14.5   &$247\pm20$\tablefootmark{$\star$}           &2.83   &$\mathcal{N}$ $(n\mathcal{N})$      &0.11   &$2.26\pm0.34$   &-20.7\\
	 354458   &J121613.0-032025    &0305800701   &3.8     &$136\pm14$\tablefootmark{$\star$}           &2.66   &$\mathcal{N}$ $(n\mathcal{N})$     &0.13   &$2.36\pm0.73$   &-21.6\\
	 356567   &J122349.5+072657   &0205090101   &21.5   &$1288\pm39$                                             &1.70   &$\mathcal{N}1$ $(n\mathcal{B})$   &0.07   &$1.82\pm0.33$   &-21.6\\
	 358675   &J121814.8+142601   &0147610101   &13.4   &$456\pm24$                                               &2.87   &$\mathcal{B}$ $(n\mathcal{B})$     &0.13   &$4.45\pm0.76$   &-20.1\\
	 369702   &J125821.3+013158   &0658400601   &14.0   &$2373\pm52$                                             &1.45   &$\mathcal{B}$ $(n\mathcal{B})$     &0.16   &$2.51\pm0.64$   &-21.1\\
	\hline
\end{tabular}
		\tablefoot{\tablefoottext{$\ast$}{This source is not an XMM-Newton target, nevertheless, 
		detailed analysis of its serendipitous XMM data is presented by \citet{Zoghbi08}  (see also 
		Upper panel of Fig.\ref{spec}).}
		\tablefoottext{1}{The sub-sample names $\mathcal{B}$, $\mathcal{N}$ and $\mathcal{N}
		1$ correspond respectively to the \emph{broad}, \emph{narrow} and \emph{unobscured 
		narrow} sub-samples. The classification given in brackets is discussed later, in section 3.4.}
		\tablefoottext{2}{The errors on the redshift $z$ and the absolute r magnitude $\mathrm{M_{r}}$ 
		are not shown because they are smaller than 1\%.}
		\tablefoottext{$\diamond$}{EPIC-MOS counts; because there is no pn data for 
		these sources.}
		\tablefoottext{$\star$}{Total EPIC counts.}}
	\end{table*}

	\subsection{X-ray spectral analysis}
	\label{subsec:xspec}
	
\begin{table*}
		\renewcommand{\arraystretch}{1.167}
		\centering
		\caption{X-ray spectral analysis results.}
		\label{TabRes}

\begin{tabular}{r c c c c c c c c l}
	\hline \hline
	\multirow{2}*{SRCID} &\multirow{2}*{Model} &\multirow{2}*{$\Gamma$} &\multirow{2}*{kT (keV)} &\multirow{2}*{$N_{H}\,(\cdot 10^{22}\,\mathrm{cm^{-2}})$\tablefootmark{(1)}} &\multirow{2}*{Red-$\chi^2$} &\multirow{2}*{goodness} &\multirow{2}*{P(ftest)} &\multirow{2}*{FeK} &\multirow{2}*{Sub-sample\tablefootmark{(2)}}\\
	 & & & & & & & \\ \hline
	 
	7207                                        &(2)   &$2.40_{-0.41}^{+0.45}$   &$0.12_{-0.04}^{+0.08}$         &$0.12_{-0.11}^{+0.21}$   &1.218                                      &31\%  &0.019   &$\checkmark$   &$\mathcal{B}$ $(n\mathcal{B})$\\
	18567                                      &(2)   &$2.19_{-0.26}^{+0.27}$   &$0.10_{-0.02}^{+0.02}$         &<0.01                              &1.176                                      &44\%  &0.004   &                         &$\mathcal{B}$ $(n\mathcal{B})$\\
	18747                                      &(1)   &$2.21_{-0.22}^{+0.25}$   &-                                             &<0.01                              &0.608                                      &8\%    &-          &                         &$\mathcal{B}$ $(n\mathcal{B})$\\
	28357                                      &(2)   &$2.38_{-0.05}^{+0.05}$   &$0.10_{-0.01}^{+0.01}$         &<0.01                              &1.013                                     &15\%   &0.006   &$\checkmark$   &$\mathcal{B}$ $(n\mathcal{B})$\\
	29911                                      &(2)   &$1.67_{-0.79}^{+0.58}$   &$0.22_{-0.06}^{+0.09}$         &<0.01                              &0.853                                     &15\%   &0.298   &$\checkmark$   &$\mathcal{B}$ $(n\mathcal{B})$\\
	93560\tablefootmark{$\ast$}   &(1)   &$2.51_{-0.54}^{+0.64}$   &-                                             &$0.25_{-0.16}^{+0.20}$  &1.283\tablefootmark{$\bullet$}   &-      &-          &                         &$\mathcal{B}$ $(n\mathcal{B})$\\
	115294\tablefootmark{$\ast$} &(1)   &$2.17_{-0.30}^{+0.34}$   &-                                             &<0.01                              &0.840\tablefootmark{$\bullet$}  &-      &-           &                         &$\mathcal{B}$ $(n\mathcal{B})$\\
	126294                                    &(2)   &$2.30_{-0.28}^{+0.29}$   &$0.038_{-0.033}^{+0.0037}$ &<0.01                              &1.14                                        &26\%  &0.212   &                         &$\mathcal{B}$ $(n\mathcal{B})$\\
	155872                                    &(2)   &$2.03_{-0.43}^{+0.48}$   &$0.15_{-0.04}^{+0.02}$         &<0.01                              &1.106                                     &35\%   &0.004   &$\checkmark$   &$\mathcal{B}$ $(n\mathcal{B})$\\
	282027\tablefootmark{$\ast$} &(1)   &$2.40_{-0.46}^{+0.59}$   &-                                             &<0.01                              &1.332\tablefootmark{$\bullet$}  &-      &-           &                         &$\mathcal{B}$ $(n\mathcal{B})$\\
	294033                                    &(2)   &$2.03_{-0.53}^{+0.58}$   &$0.13_{-0.02}^{+0.02}$         &<0.01                              &0.914                                     &15\%   &0.052   &                         &$\mathcal{B}$ $(n\mathcal{B})$\\	
	297574                                    &(2)   &$2.03_{-0.35}^{+0.36}$   &$0.13_{-0.03}^{+0.02}$         &<0.01                              &0.848                                     &13\%   &0.0004 &$\checkmark$   &$\mathcal{B}$ $(n\mathcal{B})$\\
	321587                                    &(2)   &$2.21_{-0.13}^{+0.13}$   &$0.12_{-0.02}^{+0.02}$         &<0.01                              &1.033                                      &24\%  &0.014   &$\checkmark$   &$\mathcal{B}$ $(n\mathcal{B})$\\	
	324308                                    &(1)   &$1.98_{-0.10}^{+0.11}$   &-                                             &<0.01                              &1.064                                     &34\%   &-           &                         &$\mathcal{B}$ $(n\mathcal{B})$\\
	341733                                    &(2)   &$1.88_{-0.22}^{+0.26}$   &$0.14_{-0.06}^{+0.03}$         &<0.01                              &0.905                                     &14\%   &0.051   &                         &$\mathcal{B}$ $(n\mathcal{B})$\\
	350309                                    &(2)   &$2.19_{-0.22}^{+0.23}$   &$0.09_{-0.02}^{+0.02}$         &<0.01                              &1.008                                      &24\%   &0.0003 &                        &$\mathcal{B}$ $(n\mathcal{B})$\\
	358675                                    &(2)   &$1.66_{-0.20}^{+0.20}$   &$0.04_{-0.02}^{+0.02}$         &<0.01                              &1.395                                     &53\%   &0.029   &                         &$\mathcal{B}$ $(n\mathcal{B})$\\
	36405\tablefootmark{$\ast$}   &(1)   &$2.29_{-0.26}^{+0.28}$   &-                                             &<0.01                              &0.770\tablefootmark{$\bullet$}  &-      &-           &                         &$\mathcal{B}$ $(n\mathcal{B})$\\	
	369702                                    &(2)   &$2.19_{-0.42}^{+0.46}$   &$0.12_{-0.06}^{+0.05}$         &<0.01                              &1.033                                      &18\%  &$1.6\cdot10^{-6}$ &        &$\mathcal{B}$ $(n\mathcal{B})$\\
	3161\tablefootmark{$\ast$}     &(1)   &$0.92_{-0.21}^{+0.39}$   &-                                             &<0.14                              &0.912\tablefootmark{$\bullet$}  &-      &-           &                         &$\mathcal{N}$ $(n\mathcal{N})$\\
	3173                                        &(1)   &$1.74_{-0.37}^{+0.43}$   &-                                             &$0.41_{-0.18}^{+0.24}$  &1.056                                      &25\%  &-          &$\checkmark$   &$\mathcal{N}$ $(n\mathcal{N})$\\
	120992\tablefootmark{$\ast$} &(1)   &$1.99_{-1.31}^{+1.75}$   &-                                             &$7.11_{-4.42}^{+7.29}$  &0.663\tablefootmark{$\bullet$}  &-	&-          &$\checkmark$   &$\mathcal{N}$ $(n\mathcal{N})$\\
	155684\tablefootmark{$\ast$} &(1)   &$1.54_{-0.63}^{+1.18}$   &-                                             &<1.81                              &1.119\tablefootmark{$\bullet$}  &-      &-           &                         &$\mathcal{N}$ $(n\mathcal{N})$\\
	266064\tablefootmark{$\ast$} &(1)   &$1.22_{-1.17}^{+1.51}$   &-                                             &$23.3_{-13.8}^{+22.8}$  &0.971\tablefootmark{$\bullet$}  &-      &-           &                         &$\mathcal{N}$ $(n\mathcal{N})$\\
	277671\tablefootmark{$\ast$} &(1)   &$0.79_{-0.25}^{+0.25}$   &-                                             &<0.11                              &0.709\tablefootmark{$\bullet$}  &-      &-           &                         &$\mathcal{N}$ $(n\mathcal{N})$\\
	348033\tablefootmark{$\ast$} &(1)   &$1.62_{-0.70}^{+1.64}$   &-                                             &<1.61                              &0.701\tablefootmark{$\bullet$}  &-      &-           &                         &$\mathcal{N}$ $(n\mathcal{N})$\\	
	350410\tablefootmark{$\ast$} &(1)   &$0.97_{-0.42}^{+0.57}$   &-                                             &$0.42_{-0.33}^{+0.69}$  &0.723\tablefootmark{$\bullet$} &-	&-          &$\checkmark$   &$\mathcal{N}$ $(n\mathcal{N})$\\
	354458\tablefootmark{$\ast$} &(1)   &$1.87_{-0.39}^{+0.43}$   &-                                             &<0.04                              &1.008\tablefootmark{$\bullet$}  &-      &-           &                         &$\mathcal{N}$ $(n\mathcal{N})$\\
	1780                                        &(1)   &$2.47_{-0.45}^{+0.51}$   &-                                             &<0.11                              &0.914                                     &33\%   &-           &                         &$\mathcal{N}1$ $(n\mathcal{N}1)$\\
	28899                                      &(1)   &$2.92_{-0.23}^{+0.26}$   &-                                             &<0.01                              &0.536                                     &4\%     &-          &                         &$\mathcal{N}1$ $(n\mathcal{B})$\\
	47793\tablefootmark{$\ast$}   &(1)   &$1.96_{-0.36}^{+0.43}$   &-                                             &<0.08                              &0.669\tablefootmark{$\bullet$}  &-      &-           &                         &$\mathcal{N}1$ $(n\mathcal{N}1)$\\
	88476                                      &(2)   &$1.94_{-0.30}^{+0.30}$   &$0.07_{-0.03}^{+0.04}$         &<0.01                              &0.508                                      &1\%    &0.013   &                         &$\mathcal{N}1$ $(n\mathcal{B})$\\
	93640                                      &(2)   &$2.24_{-0.21}^{+0.22}$   &$0.016_{-0.006}^{+0.078}$   &<0.05                              &0.764\tablefootmark{$\bullet$}  &-      &0.012   &                         &$\mathcal{N}1$ $(n\mathcal{N}1)$\\
	266125\tablefootmark{$\ast$} &(1)   &$2.07_{-0.47}^{+0.52}$   &-                                             &<0.26                              &1.006\tablefootmark{$\bullet$}  &-      &-           &                         &$\mathcal{N}1$ $(n\mathcal{N}1)$\\
	275370\tablefootmark{$\ast$} &(1)   &$6.90_{-4.16}^{+}$          &-                                             &<0.01                              &0.577\tablefootmark{$\bullet$}  &-      &-           &                         &$\mathcal{N}1$ $(n\mathcal{N}1)$\\
	291676\tablefootmark{$\ast$} &(1)   &$3.02_{-0.36}^{+0.63}$   &-                                             &<0.09                              &0.958\tablefootmark{$\bullet$}  &-	&-          &                         &$\mathcal{N}1$ $(n\mathcal{B})$\\
	303293\tablefootmark{$\ast$} &(1)   &$1.71_{-0.38}^{+0.46}$   &-                                             &<0.01                              &1.029\tablefootmark{$\bullet$}  &-	&-          &                         &$\mathcal{N}1$ $(n\mathcal{N}1)$\\
	305003                                    &(1)   &$2.33_{-0.11}^{+0.12}$   &-                                             &<0.03                              &0.839                                      &4\%    &-          &                         &$\mathcal{N}1$ $(n\mathcal{N}1)$\\
	337999                                    &(1)   &$1.83_{-0.25}^{+0.28}$   &-                                             &$0.19_{-0.09}^{+0.11}$  &1.241                                      &47\%  &-           &                         &$\mathcal{N}1$ $(n\mathcal{B})$\\
	339379\tablefootmark{$\ast$} &(1)   &$1.07_{-0.38}^{+0.40}$   &-                                             &<0.01                              &1.089\tablefootmark{$\bullet$}  &-      &-          &                         &$\mathcal{N}1$ $(n\mathcal{N}1)$\\
	356567                                    &(1)   &$2.66_{-0.12}^{+0.13}$   &-                                             &<0.01                              &0.826                                     &13\%   &-          &                         &$\mathcal{N}1$ $(n\mathcal{B})$\\
	\hline
\end{tabular}
	
		\tablefoot{All quoted errors are for a 90\% confidence interval for one parameter.
		\tablefoottext{1}{The lower limit of 0.1 on $N_H$ corresponds to the limit imposed by 
		XSPEC.}
		\tablefoottext{2}{The sub-sample names $\mathcal{B}$, $\mathcal{N}$ and $\mathcal{N}
		1$ correspond respectively to the \emph{broad}, \emph{narrow} and \emph{unobscured 
		narrow} sub-samples. The classification given in brackets is discussed later, in section 3.4.}
		\tablefoottext{$\ast$}{Very low number of counts (Cash statistic is used instead of 
		$\chi^2$ statistic), not sufficient to check the presence of a soft excess.}
		\tablefoottext{$\bullet$}{Test statistic value when the Cash fit statistic is used.}
		Model (1) corresponds to an absorbed power-law (\texttt{tbabs*(zpo*pha)}) and model 
		(2) to a soft excess in addition to an absorbed power-law (\texttt{tbabs*((zpo+zbb)*pha)}).
		}
	\end{table*}

	An X-ray spectral analysis of the elusive AGN has been carried out based on the 
	XMM-Newton EPIC data and using XSPEC (version 12.8.0). The source parameters and 
	classification are given Table \ref{TabParam}. 
		
	For the sources with more than 500 counts in the three EPIC cameras combined, spectra are 
	already extracted as pipeline data products associated with the 3XMM catalogue. In this case, we 
	use the EPIC-pn spectra from the PPS files with the corresponding latest version of the calibration 
	files (SAS 11.0).
	
	For the other sources, the raw observation data files (ODF) were reduced and analysed using 
	the Science Analysis System (SAS) tool (xmmselect version 2.65.2) using standard techniques. 
	The spectral data were extracted from an optimised circular region, and the corresponding 
	background files from a nearby circular region free of sources. Because of the low number of 
	counts, to improve the analysis, the EPIC-pn and EPIC-MOS spectra are analysed simultaneously 
	with a constant coefficient inserted to model calibration uncertainties between the detectors 
	(coefficient fixed to 1 for the pn and allowed to vary for the MOS data). In the few cases for which 
	pn spectra are not available, we use the MOS data. 
	
	The spectral data are grouped with a minimum of 20 counts per bin in order to use the $\chi^2$ 
	fit statistics. However, for the lowest quality spectra, data are grouped with only one count per 
	bin and, in this case, we use Cash statistics \citep{Cash79}.
	
	\begin{figure}
	   \centering
	   \includegraphics[width=\hsize, trim=1.5cm 1.4cm 2.8cm 2.8cm,clip]{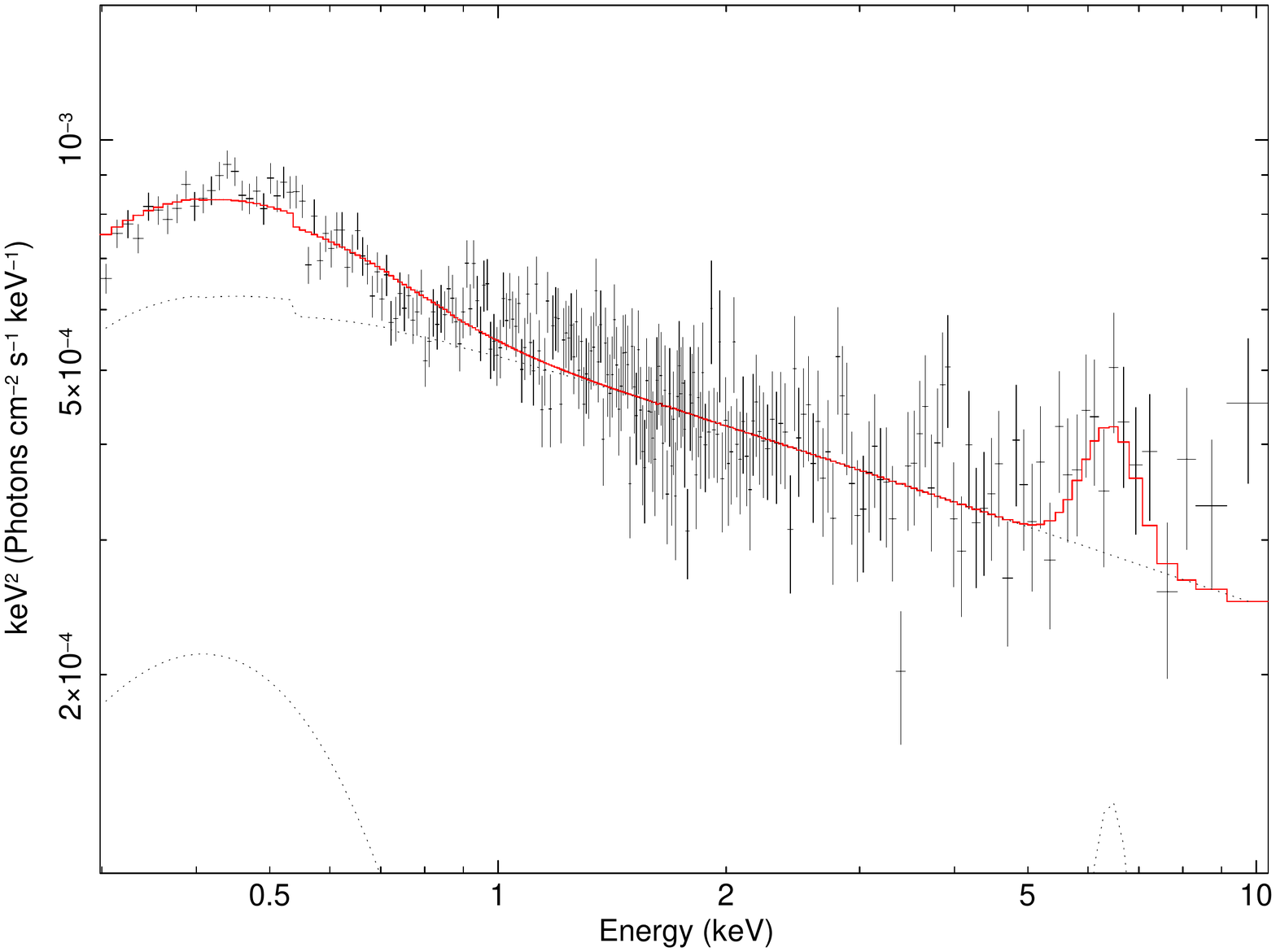}
	   \includegraphics[width=\hsize, trim=1.5cm 1.4cm 2.8cm 2.8cm,clip]{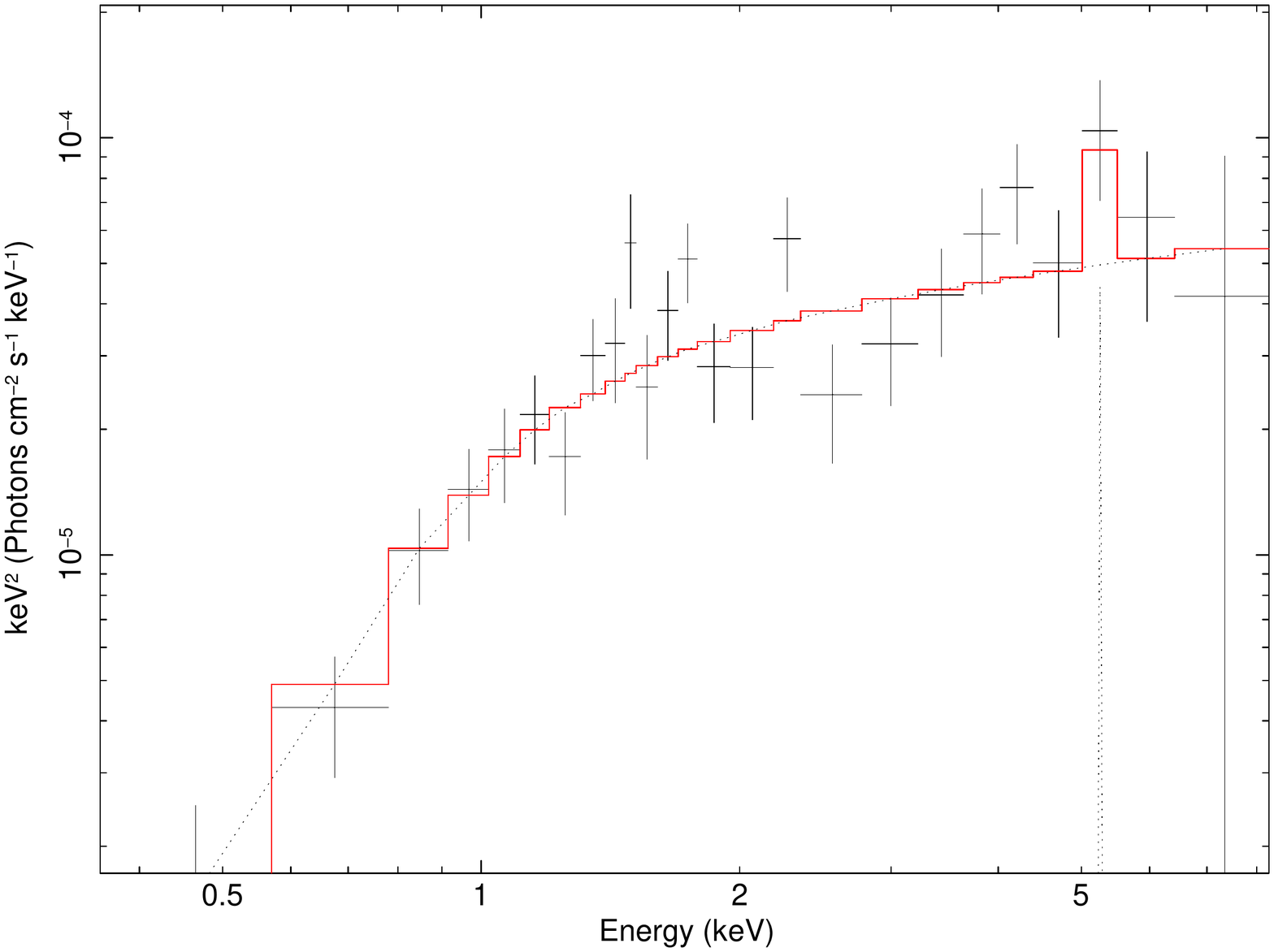}
	   \caption{Example X-ray spectra. \textbf{Upper panel}: NLS1 candidate (SRCID 28357) fitted 
	   with a power-law plus a blackbody component for the soft excess and a Gaussian component 
	   for the Fe line. \textbf{Lower panel}: a \emph{narrow} source (SRCID 3173) fitted with a 
	   Gaussian component (Fe line) in addition to an absorbed power law. The dashed lines represent 
	   the separate model components while the red line is the resulting best fit model.}
	   \label{spec}
	\end{figure}
		
	The classical model of the X-ray spectrum of an AGN is a power-law modified by absorption, with 
	in some cases a soft X-ray excess (usually present in NLS1 spectra) and a Fe K$\alpha$ 
	emission line around 6.4 keV. The absorbed power-law comes from the Comptonization of 
	photons from the accretion disc by a corona of hot material. The soft X-ray excess may be 
	attributed to the blurred ionized reflection from the inner parts of the accretion disc 
	\citep{Gier04,Crummy06} or to the Comptonization of extreme UV accretion disc photons 
	\citep{Ross92}.  The Fe K$\alpha$ fluorescent emission-line is a signature of X-ray 
	reprocessing (reflection of the power-law component by a cold disc; accretion disc or torus).\\
		
	The spectral fitting is performed as follows: 
	\begin{enumerate}
	    \item We first fit the data in the 2-10 keV energy range with a power-law (\texttt{zpo} model in 
	    XSPEC syntax) corrected for foreground absorption in our own Galaxy (\texttt{tbabs}; 
	    $N_{H,gal}$) and modified by intrinsic absorption (\texttt{pha}; $N_H$). The $N_{H,gal}$ 
	    parameter is fixed to the value computed using the emph{nh ftool} (see Table 
	    \ref{TabParam}), while $N_H$ is free to vary.
	    \item We then check for the presence of an additional Fe emission line around 6.4 keV, which 
	    can be modelled by a component with a Gaussian profile (\texttt{zgauss}).
	    \item We finally include the data between 0.3 and 2 keV to see if these data are in excess 
	    compared to the previously fitted power-law. If this is the case, we add a blackbody 
	    component (\texttt{zbbody}) to model the soft excess.
	\end{enumerate}
	
	\begin{figure}
	   \centering
	   \includegraphics[width=\hsize]{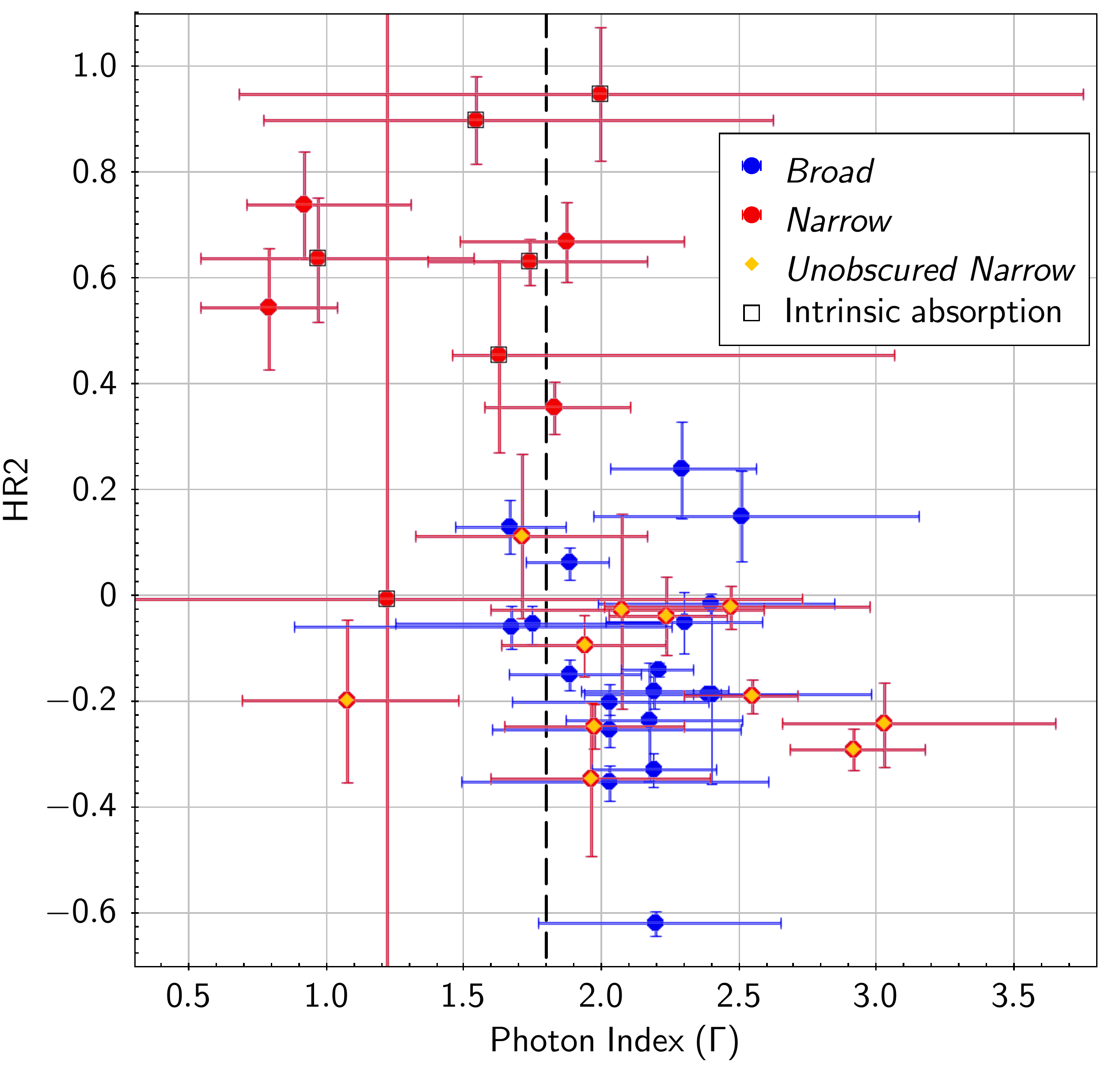}
	   \caption{HR2 vs. X-ray photon index. NLS1 candidates (in blue) have $\Gamma$ mostly larger 
	   than 1.8 and are unabsorbed, as expected. For narrow-line sources (in red), some have typical 
	   properties of Seyfert 2 ($\Gamma  < 1.8$ and intrinsic absorption) while the others, with low 
	   HR2, cannot be clearly classified using their photon index. One of the sources with very 
	   large photon index value and error is not plotted.}
	   \label{HRvsPhoInd}
	\end{figure}	
	
	The best fit model is the one that gives the smallest reduced-$\chi^2$ ($=\chi^2/\mathrm{d.o.f}$). 
	When the best fit model is the combination of 2 components (i.e. an absorbed power-law plus a 
	soft excess and/or a Fe line), we use the XSPEC \emph{Ftest} task to determine whether the 
	inclusion of an additional component is needed. We also check the validity of the model using 
	the XSPEC \emph{goodness} parameter which returns the percentage of simulations for which 
	the statistic is less than that for the observed data; so a smaller \emph{goodness} parameter 
	values indicates a better model (but this can only be used for $\chi^2$ statistic). We choose to 
	keep the additional component if the F-test probability $P(ftest)<10^{-2}$, or if the 
	\emph{goodness} parameter is really improved (i.e. a parameter difference of at least 20\%).
	
	The results of the spectral fitting are given Table \ref{TabRes}. In the case of very low quality 
	spectra, the number of counts was not sufficient to check the presence of a soft excess. Two 
	example spectra (one for a source from the emph{broad} sub-sample and one from the 
	\emph{narrow} sub-sample) are shown in Fig.\ref{spec}.\\	
	A Seyfert galaxy is expected to have a photon index of about 1.8 \citep{Nandra97}, while NLS1 
	have a softer X-ray slope, steeper than that of classical Seyferts \citep{Botte04}. The sources 
	from the \emph{broad} sub-sample (i.e. NLS1 candidates) have a photon index $\Gamma$ 
	consistent with what is expected for NLS1 ($\Gamma>1.8$) and have no or little intrinsic 
	absorption ($N_H<4\cdot10^{21}\,\mathrm{cm^{-2}}$ is often taken to be the separation 
	between absorbed and unabsorbed AGNs; \citet{Cac07}); strongly supporting their classification. The 
	sources from the \emph{narrow} sub-sample, with high HR2 (>0.4), have $\Gamma<1.8$ and show 
	signs of absorption in their spectra, so their X-ray emission resemble that of classical Seyfert 2 
	(Sy2) galaxies; while the others (\emph{unobscured narrow} sub-sample) have a $\Gamma 
	\gtrsim 1.8$ and their X-ray spectra alone cannot distinguish between NLS1 and Sy2 (see Fig. 
	\ref{HRvsPhoInd}). The X-ray spectral analysis nevertheless confirms that the sample sources with a 
	low HR do not show signs of absorption or have little intrinsic absorption.
			
	\subsection{Forbidden vs. permitted lines width}
	\label{sec:ForbPerm}

\begin{figure}
	   \centering
	   \includegraphics[width=\hsize]{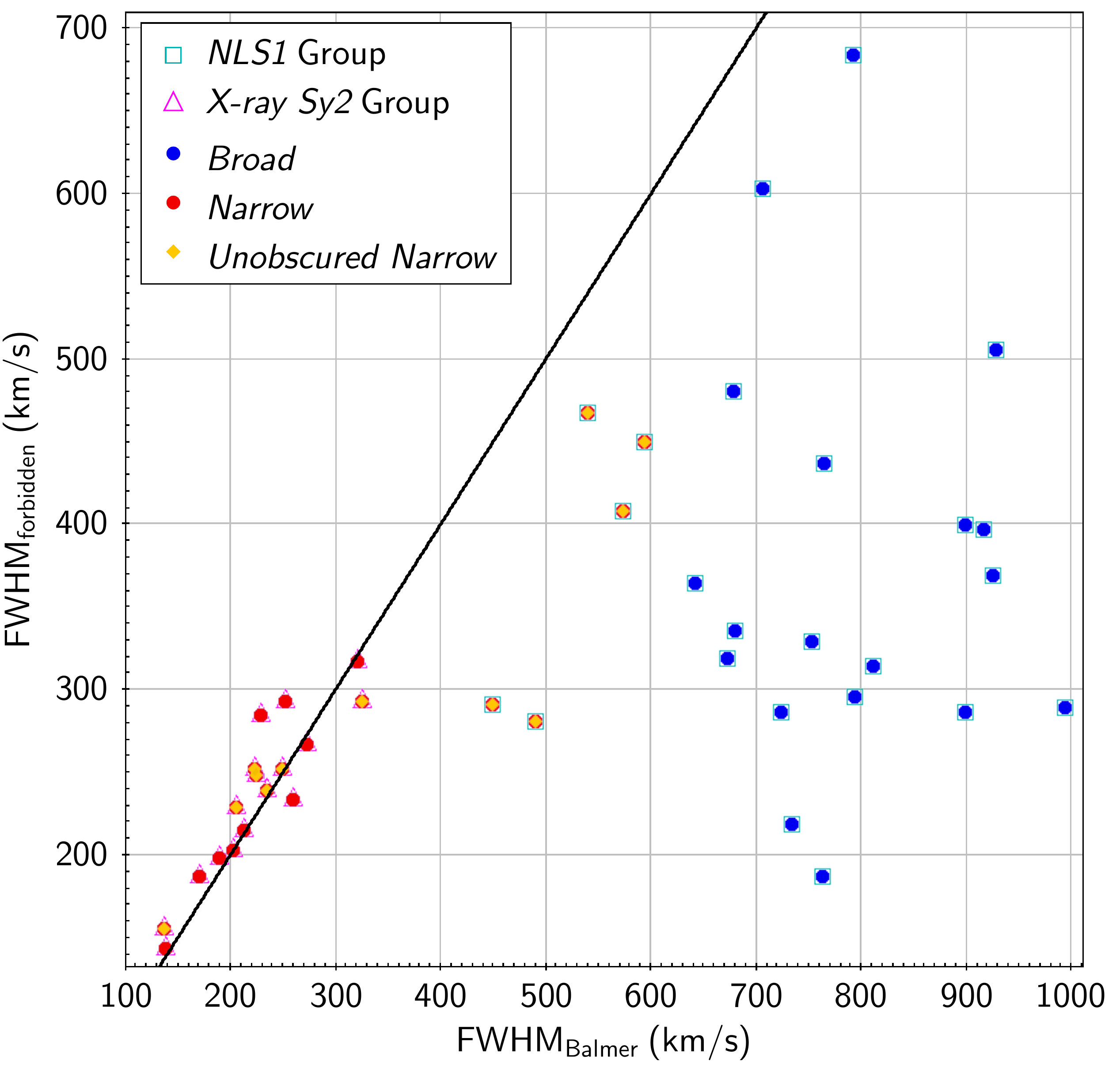}
	   \caption{Forbidden vs. permitted line widths. The black line corresponds to equal width. This 
	   comparison is used to distinguish between \emph{X-ray Sy2} (sources with equal line width) 
	   and \emph{NLS1} (which consists of all the \emph{broad} sources (blue symbols) plus 5 
	   sources from the \emph{unobscured narrow} group).}
	   \label{FWHMfvsb}
	\end{figure}

	To better distinguish between Sy2 and NLS1 (especially for the sources from the 
	\emph{unobscured narrow} group with $\Gamma \sim 1.8$), we examine one of the NLS1 
	properties, the emission line width. Because in NLS1 the broad lines come from the BLR, which is 
	presumed to be closer to the BH than the NLR, they have permitted line widths larger that for the 
	forbidden lines \citep{Botte04}. As we can see in Fig. \ref{FWHMfvsb} there are clearly two kinds of 
	sources. Some have the same line width for the permitted and forbidden lines (these we call the 
	\emph{X-ray Seyfert 2} class; 41\%) whilst others have larger permitted line widths 
	compared to their forbidden lines (\emph{NLS1} class; 59\%). The \emph{NLS1} class, 
	(referring to $n\mathcal{B}$ in Tables 1 and 2), consists of all the sources from the 
	\emph{broad} sub-sample (i.e. NLS1 candidates) plus the five sources of the 
	\emph{unobscured narrow} sub-sample which have Balmer velocity line width not dominated 
	by the host galaxy. The \emph{X-ray Sy2} sources (named as $n\mathcal{N}$ in 
	Tables 1 and 2) are the new \emph{narrow} sources, with 8 of them having a low HR 
	($n\mathcal{N}1$ classification in Tables 1 and 2, hereafter called \emph{unobscured X-ray 
	Sy2}) and so seem to be unabsorbed.
	
	\subsection{NLS1 class: Black hole mass and Eddington ratio}
	\label{sec:NLS1}
	To confirm the classification of the NLS1 sources, we compute their BH masses and Eddington ratios 
	($\lambda = L_{bol}/L_{Edd}$). NLS1 are known to be AGN with very low mass BH and must accrete 
	at near-Eddington rates ($\lambda \gtrsim 0.25$; \citep{Netzer07}) in order to maintain an AGN-like 
	luminosity.
	
	NLS1 share the same stellar velocity dispersion range as type 1 AGN in the $M_{BH}-\sigma$ 
	relation \citep{Botte04}, so we can use the same BH mass relation as for unobscured AGN to 
	estimate the mass of their central object. The mass of a BH can be computed assuming virial 
	equilibrium: $M_{BH}=R_{BLR}v^2G^{-1}$. We use the scaling relation of \citet{Xiao11}:
	
	\begin{equation}
		M_{BH}\,(\mathrm{M_{\odot}})=3.47\cdot \left[\frac{\mathbf{\lambda}L_{5100\AA}}
		{10^{44}\, \mathrm{erg.s^{-1}}}\right]^{0.519}\cdot \left[\frac{FWHM_{H\alpha}}
		{\mathrm{km.s^{-1}}}\right]^{2.06}
		\label{eq:MBH}
	\end{equation}
	In this relationship, the size of the BLR, ($R_{BLR}$), is inferred from the AGN continuum 
	luminosity at 5100$\AA$ using the revised relation from \citet{Bentz09} calibrated using $
	\mathrm{H}\beta$. Then $\mathrm{FWHM_{H\beta}}$ is converted to $\mathrm{FWHM_{H\alpha}}$ 
	according to the empirical relationship derived in \citet{Greene05}. Finally \citet{Xiao11} assume an 
	isotropic distribution of orbits with random inclinations ($v=f\cdot FWHM=\sqrt{3}/2 \cdot FWHM$).
	The luminosity $\mathbf{\lambda}L_{5100}$ is the rest-frame luminosity corrected for Galactic 
	absorption and host galaxy stellar contribution. The FWHM measurement for my sample comes from 
	the SDSS GALSPEC spectral products and corresponds to the value fitted simultaneously for all of 
	the Balmer lines.
	
	The Eddington ratio only depends on the Eddington luminosity $L_{Edd}$ and on the bolometric 
	luminosity $L_{bol}$.
	The Eddington luminosity is directly linked to the BH mass:
	\begin{equation}
		L_{Edd}=1.25\cdot 10^{38}\left(\frac{M_{BH}}{\mathrm{M_{\odot}}}\right)\,
		\mathrm{erg.s^{-1}}
		\label{eq:LEdd}
	\end{equation}
	Because our sample have low [OIII] luminosities for AGN and the MIR emission can be contaminated 
	by SF, the best choice to compute the bolometric luminosity is to use the hard X-ray luminosity 
	($L_{bol}=L_{HX}\cdot C_{HX}$). The conversion factor $C_{HX}$ depends on $L_{HX}$ 
	\citep{Vasu07} and can be estimated using the relation from \citet{Marconi04}:

	\begin{equation}
		\centering
		\log \left(\frac{L_{bol}}{L_{HX}}\right)=1.54+0.24\mathcal{L}+0.012\mathcal{L}
		^2-0.0015\mathcal{L}^3
		\label{eq:Lbol}
	\end{equation}
	where $\mathcal{L}=\log(L_{bol})-12$, and $L_{bol}$, $L_{HX}$ are in units of $L_{\odot}$.
	
\begin{figure}[h]
	   \centering
	   \includegraphics[width=\hsize]{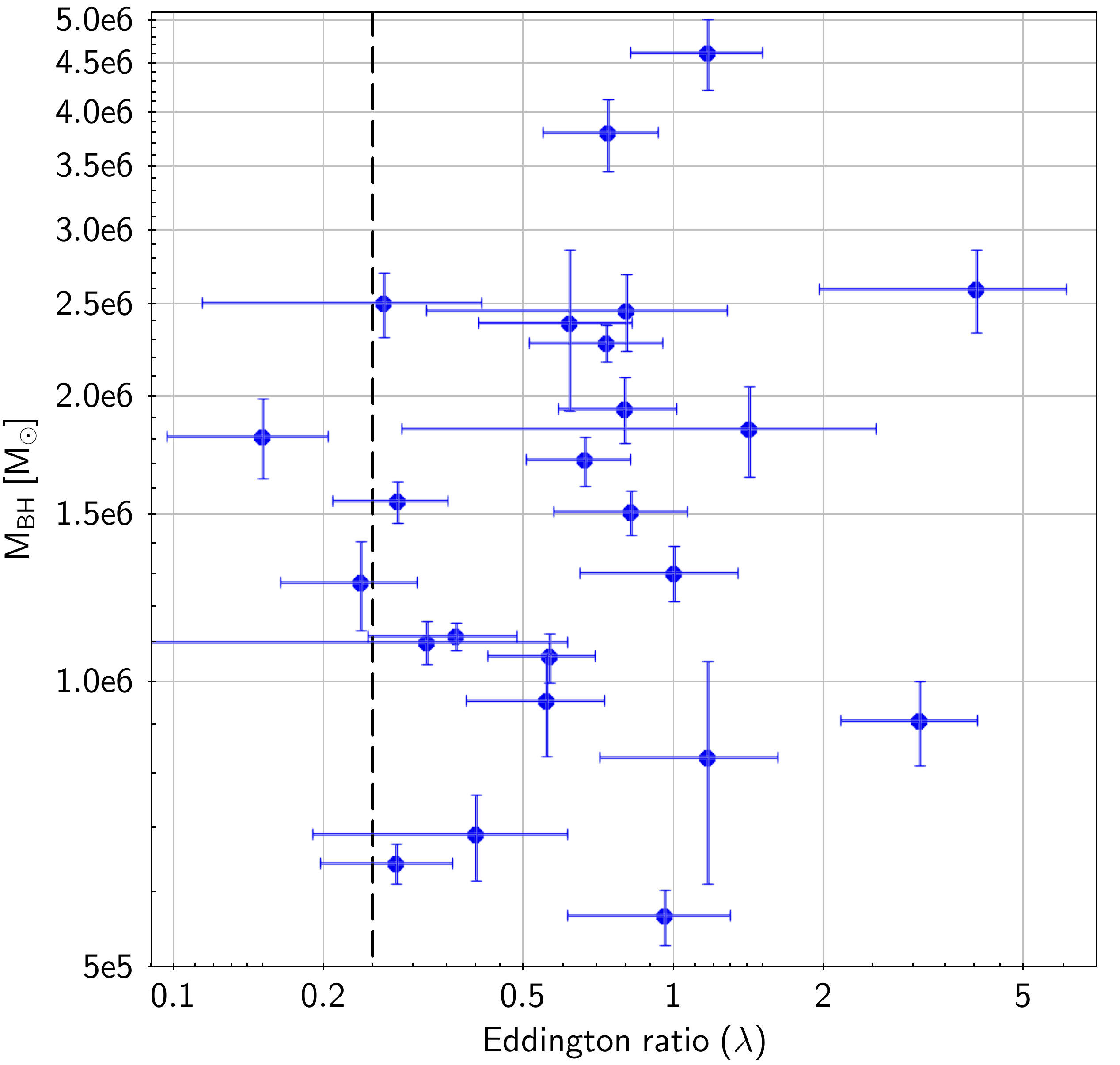}
	   \caption{$M_{BH}$ and Eddington ratio for the NLS1. The dashed line represents the lower limit
	   to the observed Eddington ratio  for NLS1 ($\lambda_{NLS1,lim}=0.25$).}
	   \label{NLS1_BH_Eddratio}
	\end{figure}

	Using equations (\ref{eq:MBH}), (\ref{eq:LEdd}) and (\ref{eq:Lbol}), the sources 
	from the \emph{NLS1} group have low mass BH ($M_{BH}<5\cdot 10^6\,\mathrm{M_{\odot}}$); 
	compared to a typical AGN which has $M_{BH}\sim 10^7-10^9\,\mathrm{M_{\odot}}$; and also 
	have  a high Eddington ratio ($\lambda \gtrsim 0.25$) as expected, as shown in Fig.
	\ref{NLS1_BH_Eddratio}.\\
	
	Among the elusive AGN,  half of sources (59\%)  are likely to be NLS1 and are optically 
	misclassified as star-forming galaxies due to the fact that Balmer lines (H$\alpha$ and H$\beta$) 
	come from both the NLR and BLR regions and therefore cannot be correctly classified using the BPT 
	diagram. However, we still have 41\% of the sample that do not appear to be NLS1, and 
	mostly resemble X-ray type 2 AGN, in the sense that they do not possess optical broad lines and 
	have AGN X-ray luminosity (in contrast with the work of \citet{Castello12} who found that the 
	population of elusive AGN is entirely comprised of NLS1). Possible reasons of their optical 
	misclassification are discussed in the next section.

\section{Why type 2 AGN could be optically misclassified as SF galaxies}
\label{sec:misclassified}
   We consider three possible explanations for the fact that the \emph{X-ray Sy2} galaxies (i.e. the 
   17 sources with $\mathrm{FWHM_{Balmer} < 400 \,km.s^{-1}}$) are not apparent as AGN in the optical 
   band: obscuration, optical dilution by host galaxy starlight and weak emission from the AGN.
	\subsection{Obscuration hypothesis}
	In order to explain the optical dullness of the sources with obscuration, we need to consider  
	complete obscuration of the central engine and/or the NLR. One possibility is that the nuclear 
	absorber is not distributed in the torus-like geometry assumed for the classical model, and so 
	obscuration may be caused by spherical Compton-thick gas clouds, covering almost 4$\pi$ at the 
	central engine \citep{Comastri02, Civano07}, implying that ionising photons cannot escape from the 
	nuclear source to produce the NLR. Or, if the covering of the nuclear source is not complete, only few 
	photons can reach the NLR, leading to attenuated emission lines from this region. On the other hand,
	\citet{Rigby06} proposed the presence of extranuclear dust and gas. These are distributed on large 
	scales (i.e. dust lanes) in the host galaxy along our line of sight and which may hide the emission 
	lines from the NLR.

	According to the X-ray spectral fitting results (see section \ref{subsec:xspec}), the hydrogen 
	column density $N_H$ is, for almost all of the sources, much smaller than $10^{22}\,
	\mathrm{cm^{-2}}$, corresponding to an optical extinction value $\mathrm{A_V\ll 4.5\,mag}$ 
	(assuming the Galactic relation: $A_V=4.5\cdot 10^{-22}\,\mathrm{N_H}$ of \citet{Guver09}), 
	and so are not heavily absorbed.
	
\begin{figure}
	   \centering
	   \includegraphics[width=\hsize]{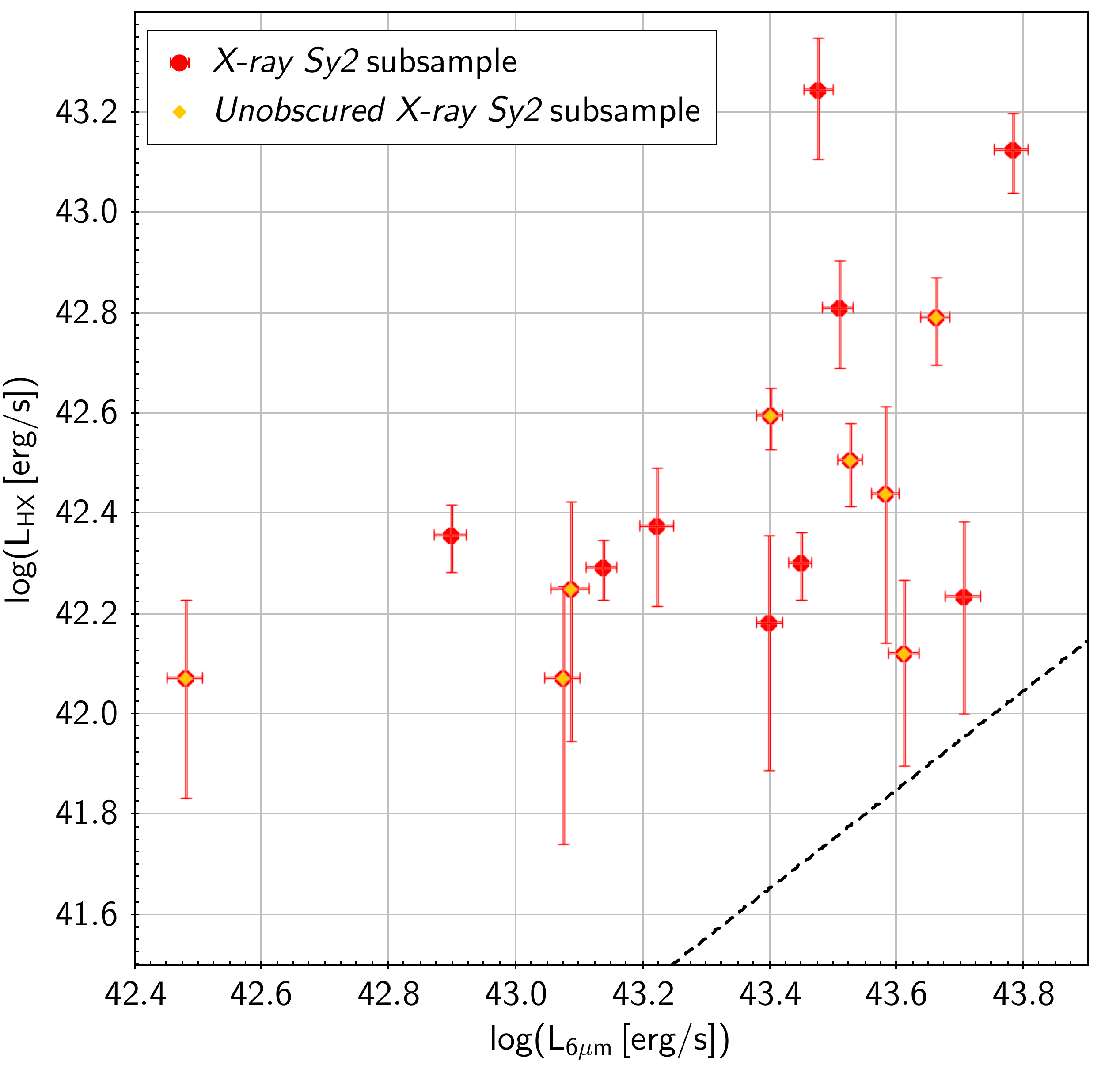}
	   \caption{Comparison of the $L_{HX}$ and $L_{6\mu m}$, two indirect AGN luminosity 
	   indicators, confirming the absence of Compton-thick sources. The Compton-thick region on the 
	   diagram lies below the black dashed line.}
	   \label{L_6um}
	\end{figure}

	For AGN with column density greater than $>10^{24}\,\mathrm{cm^{-2}}$, X-rays are significantly 
	absorbed and scattered above a few keV, so X-ray spectra below 10 keV do not provide 
	information on the real column density. The possibility that some sources are Compton-thick can 
	be investigated by evaluating the total absorption by comparing two indirect AGN luminosities. 
	The 6$\mathrm{\mu}$m continuum luminosity is a good proxy for the AGN luminosity, like the hard X-ray luminosity. We estimate the 6$\mathrm{\mu}$m flux, and thus luminosity, from the WISE fluxes in 
	the four bands (3.4$\mathrm{\mu}$m, 4.6$\mathrm{\mu}$m, 12$\mathrm{\mu}$m and 24$
	\mathrm{\mu}$m) using a weighted linear least square fitting (e.g. $\log f_\nu \propto \log 
	\lambda$). This analysis, see Fig. \ref{L_6um}, confirms that all of the sources lie above the 
	Compton-thick region of the diagram \citep{Goulding11}.
	
	We can conclude that obscuration of the central BH is unlikely to be the cause of the elusiveness of 
	optical AGN signatures in these sources. Moreover, the absence of observed obscuration in 
	the sources with narrow emission lines (i.e. \emph{unobscured X-ray Sy2}) is not due to 
	Compton-thick absorbers, which can be confirmed by comparing the [OIII] and hard X-ray 
	luminosities. In the presence of high levels of absorption, the hard X-ray luminosity is expected to be 
	depressed by an amount related to the absorbing column density, while the [OIII] emission, coming 
	from the NLR, should be unaffected. Compton-thick sources ($N_H>10^{24}\,\mathrm{cm^{-2}}$) are 
	expected to have a thickness parameter $\mathrm{T = L_{HX}/L_{[OIII]}}$  smaller than 0.1 (i.e. $\log 
	T<-1$) \citep{Bassani99}. The \emph{X-ray Sy2} class in this sample have $\log T$ between 0.5 and 
	2.5 (see Fig. \ref{L_dist}), so do not appear to be Compton-thick sources.
			
	\subsection{Optical dilution hypothesis}
	Another possible explanation for the absence of an AGN signature in the optical spectrum of 
	these sources is the optical dilution of the AGN activity by starlight from the host galaxy, with a 
	possible additional star-formation contribution.
	
	The contribution of starlight from the host galaxy can be estimated using the 4000$\AA$ 
	break ($\mathrm{\Delta_n}$):
	$$\mathrm{\Delta_n}=\frac{F^+ - F^-}{F^+}$$
	where $F^+$ and $F^-$ represent the mean value of the flux density in the rest-frame regions 
	4000-4100$\AA$ and 3850-3950$\AA$ respectively (definition from \citet{Balogh99}, less 
	sensitive to reddening by dust). The nuclear emission is dominant if the value of $
	\mathrm{\Delta_n}$ is below 20\%, whilst the host galaxy light  dominates the 
	continuum if $\mathrm{\Delta_n}$ ranges from 20\% to 60\%. For $\mathrm{\Delta_n}$ values 
	between 20\%-40\% both the AGN and the host galaxy contribute to the spectrum.

\begin{figure}
	   \centering
	   \includegraphics[width=\hsize]{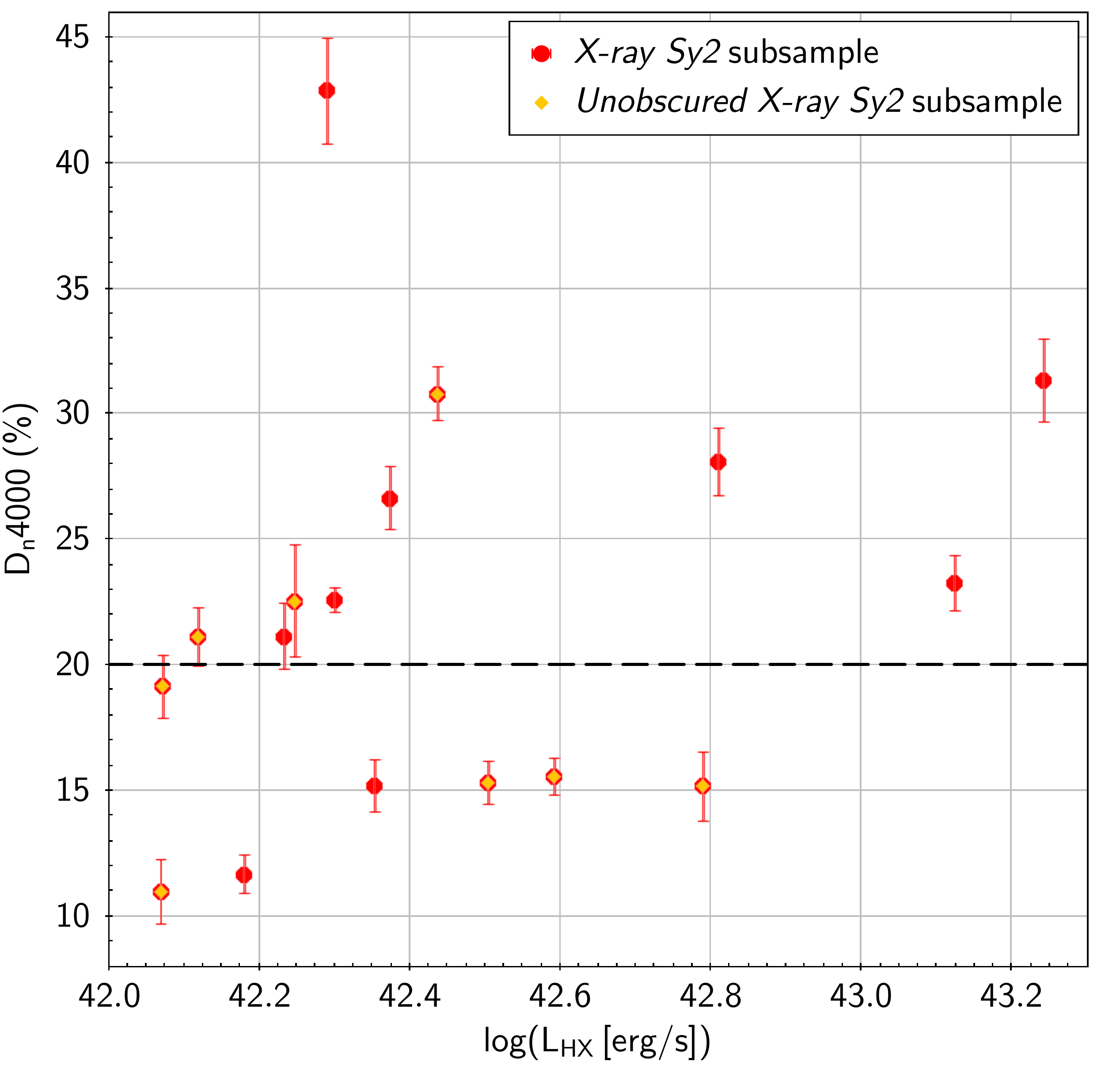}
	   \caption{4000$\AA$ break ($\mathrm{\Delta_n}$) values for the sources in the \emph{X-ray 
	   Sy2} group. The optical continuum of a galaxy is dominated by AGN for $\mathrm{\Delta_n}<20\%
	   $, and for larger values it is dominated by the host galaxy light.}
	   \label{d_4000}
	\end{figure}

	We can see in Fig. \ref{d_4000} that 10 of the 17 \emph{X-ray Sy2} sources have their blue 
	continuum dominated by the host galaxy starlight and so the AGN light is diluted.
	
	Moreover, by comparing the observed H$\alpha$ line luminosity with the value expected for an 
	AGN (using the relation from \citet{Panessa06}), most of these sources have Balmer emission 
	lines fluxes in excess, which could be due to star-formation. Any star-formation contribution will 
	enhance the Balmer lines and so reduce the $\mathrm{[OIII]/H}\beta$ and $\mathrm{[NII]/H}
	\alpha$ line ratios, moving the sources from the AGN region to the SF region of the BPT 
	diagram.
	
	So intrinsic weakness of the AGN, in respect to the host galaxy, combined in some cases with a 
	star-formation contribution could explain the elusiveness of about 60\% of the sources of the
	\emph{X-ray Sy2} group.
	
	\subsection{Low accretion rate hypothesis}
	Finally, undiluted optically dull AGN could be characterised by intrinsically weak optical emission 
	from the accretion disk. In the case of low accretion rates, AGN are expected to be optically 
	underluminous compared with typical AGN due to radiatively inefficient accretion flows (RIAFs, 
	\citet{Yuan04}).
	
	The Eddington ratio ($\lambda=L_{bol}/L_{Edd}$) of these sources is computed using the 
	equations (\ref{eq:LEdd}) and (\ref{eq:Lbol}) described in the section \ref{sec:NLS1} for the 
	bolometric luminosity. The BH masses of the Sy2 sources were estimated using the relation 
	between $\mathrm{M_{BH}}$ and the rest-frame K-band bulge luminosity (from 
	\citet{Graham07}, as improved by \citet{Trump11}):
	$$\log\left(\frac{M_{BH}}{\mathrm{M_{\odot}}}\right)=0.93\cdot\left(\log \left(L_{K,bulge}
	\right)-0.3z\right)+32.30$$
	with $L_{K,bulge}$ in $\mathrm{erg.s^{-1}}$ and $L_{K,bulge}=0.5\cdot L_{K,host}$ assumed for 
	narrow-line AGN.
			
\begin{figure}
	   \centering
	   \includegraphics[width=\hsize]{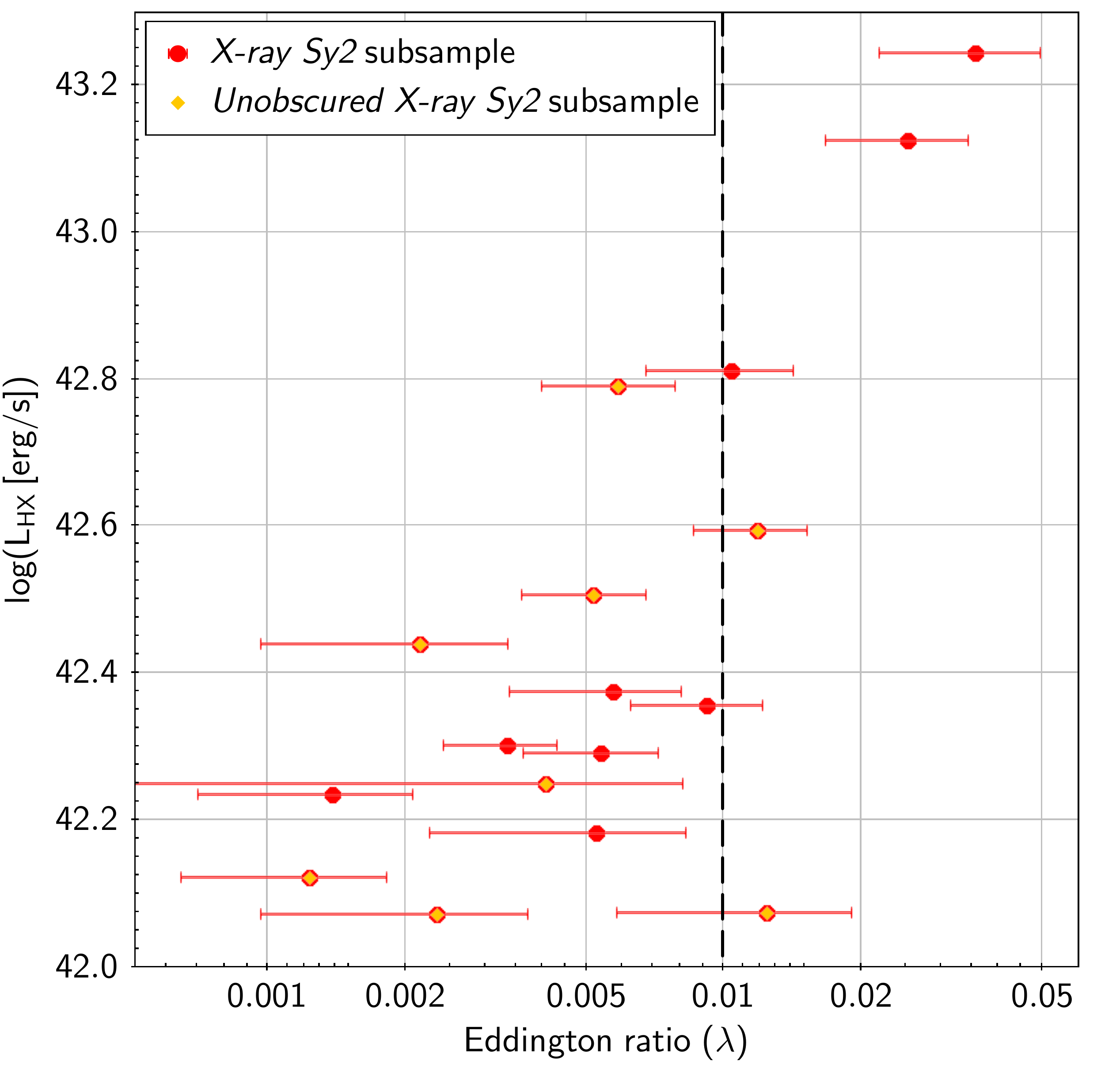}
	   \caption{Hard X-ray luminosity vs. Eddington ratio ($\lambda$) of the \emph{X-ray Sy2} 
	   sources. The values of $\lambda$ are low compared to typical AGN and the low HR2 sources (in 
	   yellow) have very low accretion rates ($\lambda<10^{-2}$; black dash line) and so are good 
	   candidates to be \emph{True Sy2}.}
	   \label{Sy2Eddratio}
	\end{figure}

	As can be seen in Fig. \ref{Sy2Eddratio}, the Sy2 sources have low Eddington ratios, $0.001<
	\lambda<0.05$, compared to typical AGN ($\lambda \sim 0.07-1$, \citet{Heckman04}), providing an 
	explanation for the optical elusiveness of these sources.Moreover, all the \emph{unobscured 
	X-ray Sy2} sources have very low accretion rates ($\lambda<10^{-2}$). A recent study by 
	\citet{Trump11} showed that for a specific accretion rates ($\lambda<10^{-2}$), the BLR disappears 
	because of the occurrence of a RIAF extending to the inner region of the accretion disk. Below this 
	specific accretion rate, the disk wind, which is supposed to form in the BLR, is no longer supported. 
	The resulting AGN are unobscured but lack broad emission lines in their optical spectrum: they are 
	called \emph{True Sy2}.\\
	Therefore, the 8 \emph{unobscured X-ray Sy2} sources are good candidates to be 
	\emph{True Sy2} AGN.
	
	\subsection{True Sy2 candidates: Expected FWHM and luminosity of the broad lines}
	\label{Sec:TrueSy2}
	
	True Sy2 candidates have no intrinsic absorption so, according to the Unified Model, are expected
	 to exhibit broad optical emission lines. To confirm their classification as True Sy2, 
	we can compute their expected BL FWHM and luminosity and compare these with the observed 
	values.
	
	For the broad H$\alpha$ line, the FWHM is directly related to the BH mass and continuum 
	luminosity at 5100$\AA$ (see equation (\ref{eq:MBH}), section \ref{sec:NLS1}). 

	\begin{figure}[h]
	   \centering
	   \includegraphics[width=\hsize]{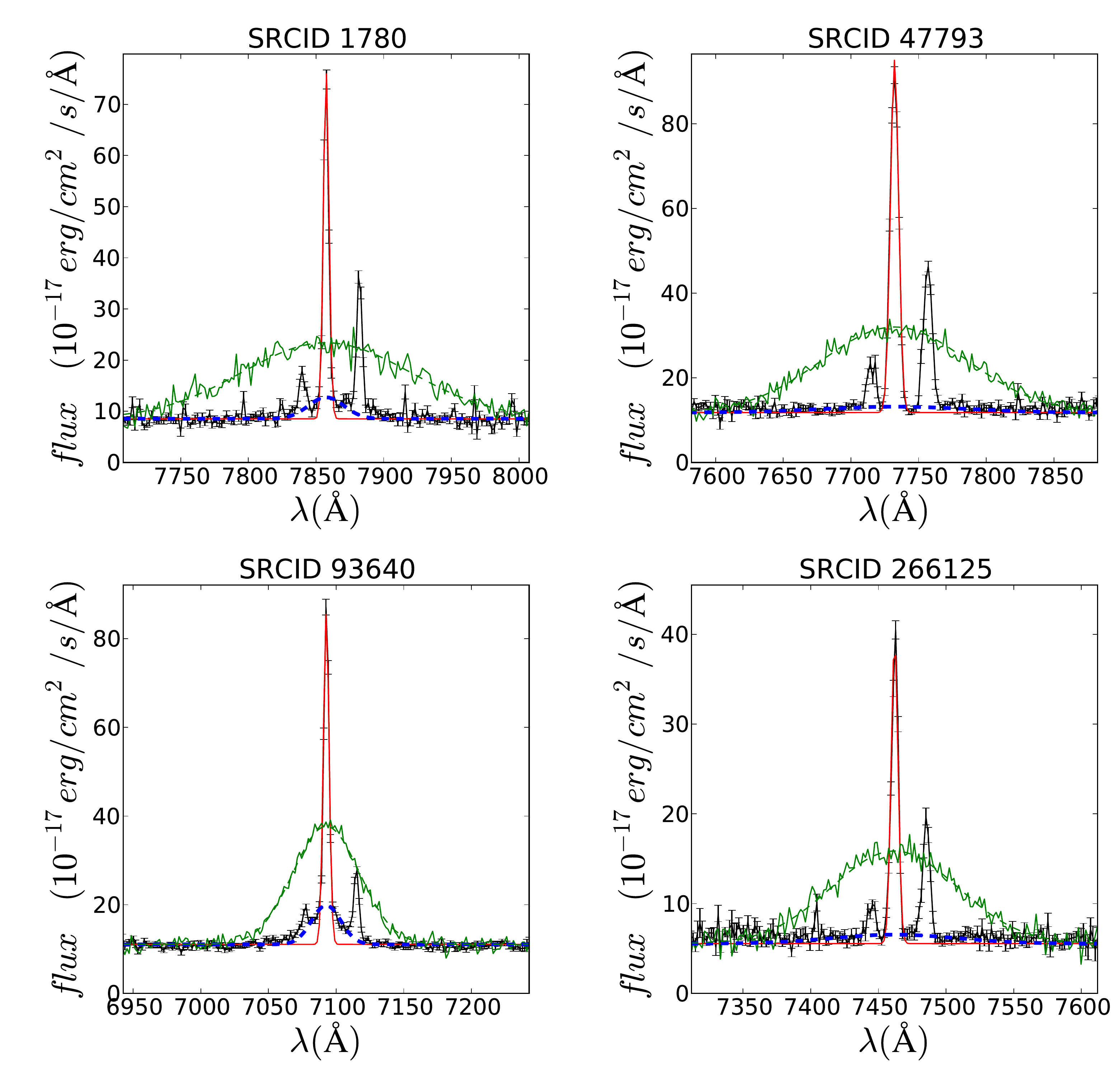}
	   \includegraphics[width=\hsize]{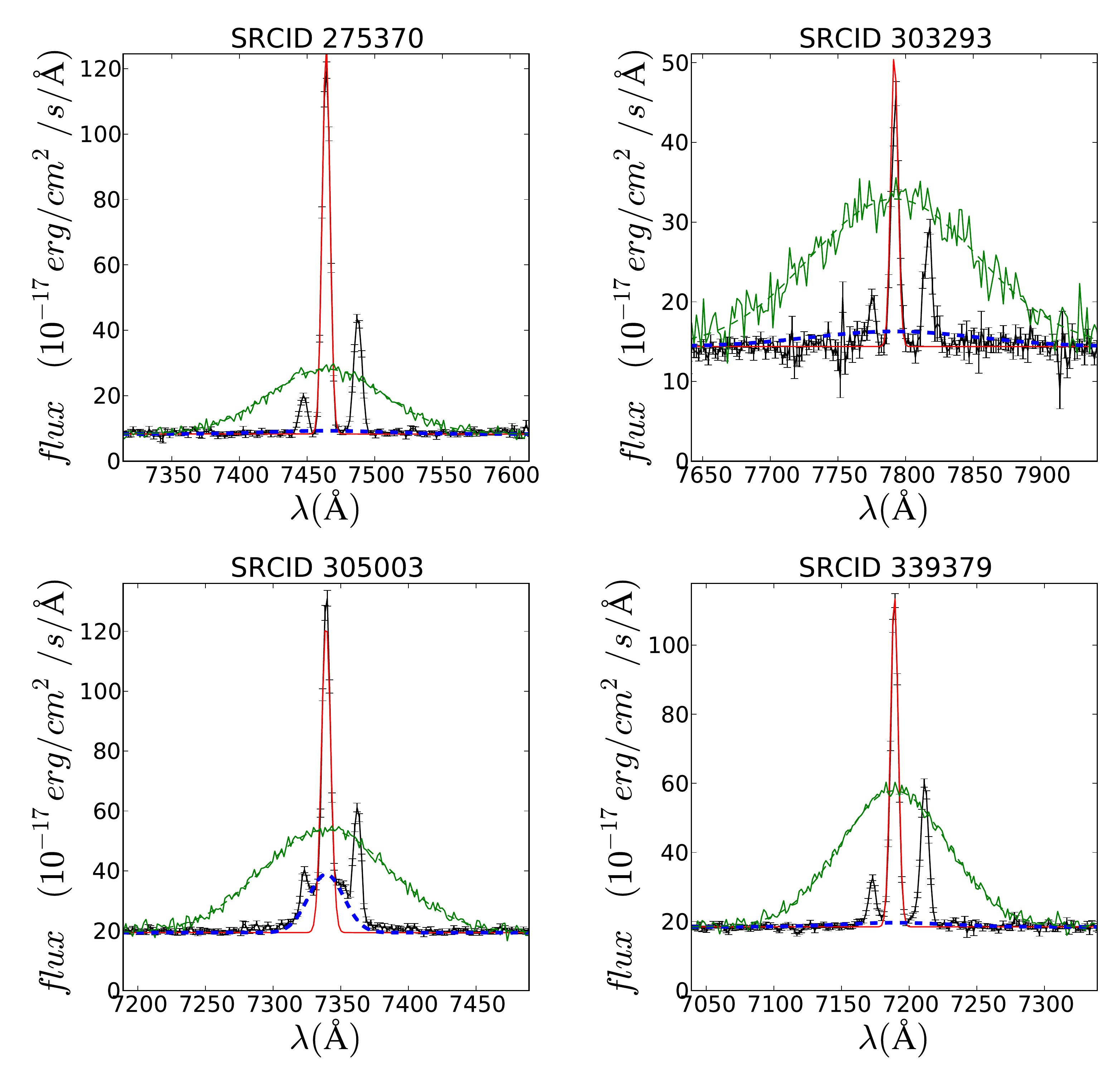}
	   \caption{Optical spectra of the True Sy2 candidates around the H$\alpha$ region. The 
	   observed H$\alpha$ line is fitted with a Gaussian profile (red curve) while the predicted 
	   broad component is represented by a Gaussian (green). The dashed blue line represents the fitted 
	   weak broad line or the upper limit to the broad component.}
	   \label{broadHa}
	\end{figure}

	The predicted broad H$\alpha$ luminosity can be estimated from the empirical relation of 
	\citet{Greene05}:
	$$L_{Broad\,H\alpha}=(5.25\pm 0.02)\cdot 10^{42}\left(\frac{\mathbf{\lambda}L_{5100}}{10^{44}\,
	\mathrm{erg.s^{-1}}}\right)^{(1.157\pm 0.005)} \,\mathrm{erg.s^{-1}}$$
	The rms scatter in this relation is quite small, being $\sim0.2$ dex.
	
	The observed flux and line width of H$\alpha$, from the optical spectrum, and the predicted values 
	for the broad component from the equations above and (\ref{eq:MBH}), are given Table 
	\ref{TabSy2Param}.
		
	The observed spectrum in the H$\alpha$ region with the predicted broad H$\alpha$ component 
	superimposed is shown in Fig. \ref{broadHa}. The H$\alpha$ line is fitted with a Gaussian 
	profile. The predicted broad component is modelled by Gaussian function with width and flux as 
	discussed above; the predicted component is plotted with randomised measurement noise 
	equivalent to that of the actual observed spectrum. The broad line flux plotted in Fig. \ref{broadHa} 
	is the fitted value (SRCIDs 1780, 93640, 305003) where a possible weak line is detected or, if no 
	line is detected, an upper limit, set to twice the apparent noise level (see Table 
	\ref{TabSy2Param}, 
	column 3).
	
	\begin{figure}[b]
	   \centering
	   \includegraphics[width=\hsize]{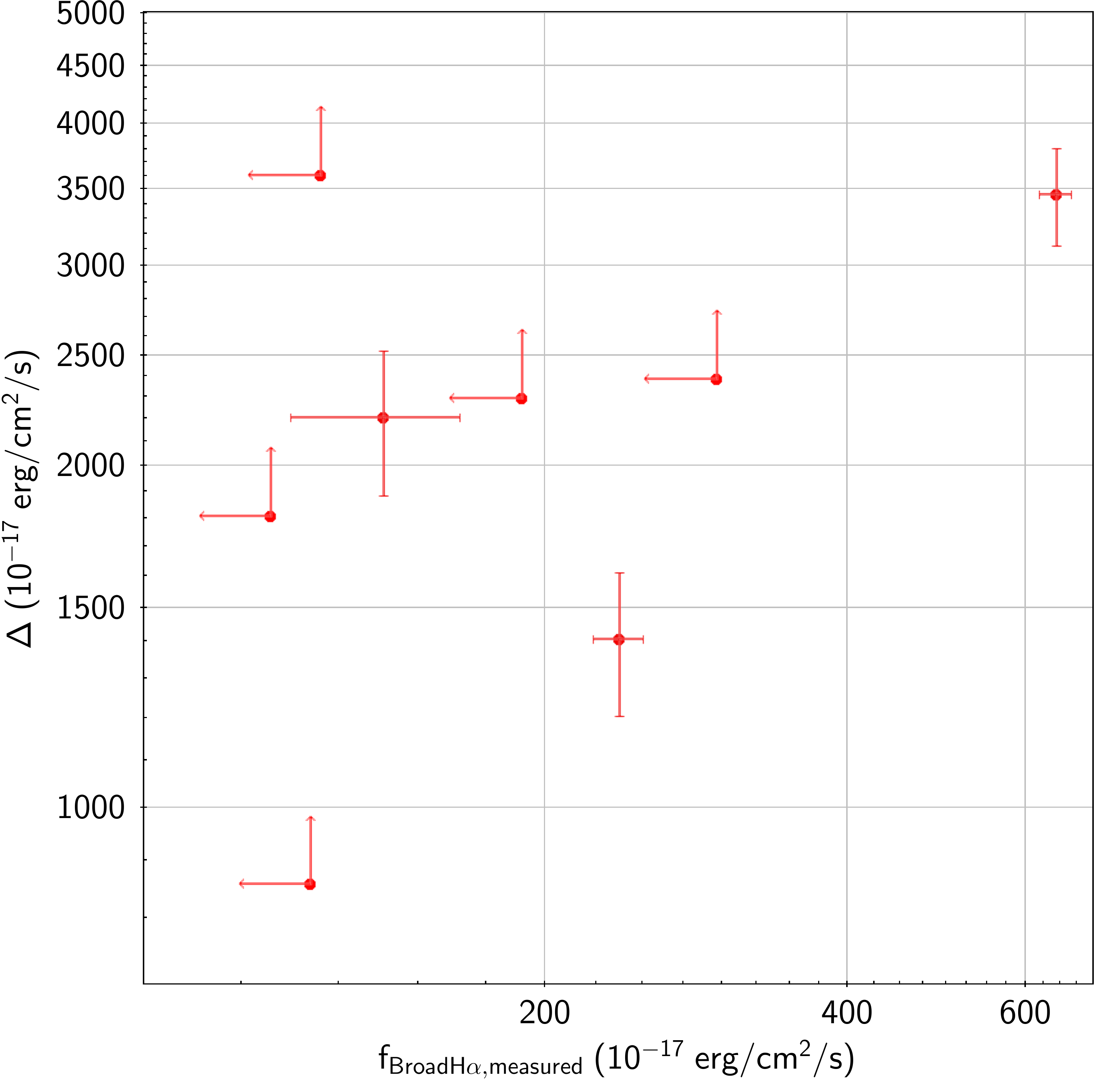}
	   \caption{Difference between the predicted and measured flux (noted as $\Delta$) vs. 
	   measured flux for the H$\alpha$ broad component of the True Sy2 candidates.}
	   \label{ObsvsMeasF}
	\end{figure}
	
	\begin{table*}   
		\centering
		\caption{True Sy2: observed and predicted parameters for H$\alpha$.}
		\label{TabSy2Param}

\begin{tabular}{c| c c| c c| c c}
	\hline \hline
	\multirow{2}*{SRCID} &\multicolumn{2}{|c|}{Narrow H$\alpha$\tablefootmark{1}} &\multicolumn{2}{c|}{Broad H$\alpha$\tablefootmark{2}}  &\multicolumn{2}{c}{Predicted Broad H$\alpha$\tablefootmark{3}} \\
	\cline{2-7}
	 &$flux$\tablefootmark{a} &$FWHM$\tablefootmark{b} &$flux$\tablefootmark{a} &$FWHM$\tablefootmark{b,$\ast$} &$flux$\tablefootmark{a} &$FWHM$\tablefootmark{b}  \\ \hline 
	 1780       &$357\pm6$     &$188\pm3$     &$139\pm26$      &$1181\pm430$    &$2338\pm295$   &$5638\pm375$     \\   
	 47793     &$631\pm8$     &$275\pm3$     &<190                  &4940                    &$2741\pm264$   &$4940\pm286$     \\
	 93640     &$398\pm5$     &$212\pm2$     &$238\pm14$      &$1071\pm278$    &$1642\pm190$   &$2454\pm279$     \\
	 266125   &$216\pm4$     &$242\pm4$     &<117                  &4428                    &$1210\pm238$   &$4428\pm1769$   \\
	 275370   &$843\pm9$     &$266\pm2$     &<107                  &4047                    &$2130\pm221$   &$4047\pm246$     \\
	 303293   &$243\pm43$   &$237\pm16$   &<296                  &5638                    &$2887\pm209$   &$5638\pm303$     \\
	 305003   &$887\pm9$     &$327\pm3$     &$644\pm23$      &$1265\pm268$    &$4099\pm318$   &$4579\pm177$     \\
	 339379   &$691\pm7$     &$280\pm2$     &<119                  &3915                    &$3934\pm222$   &$3915\pm155$     \\
	 \hline
\end{tabular}
		\tablefoot{\tablefoottext{1}{Parameters from the optical spectrum, obtained by the SDSS line 
		fitting pipeline (SDSS \emph{SpecLine} table) (Fig. \ref{broadHa} red curve).}
		\tablefoottext{2}{Fitted weak line (SRCIDs 1780, 93640, 305003) or the upper limit of the 
		broad component (Fig. \ref{broadHa} dashed blue line).}
		\tablefoottext{3}{Predicted broad H$\alpha$ component computed using equations of
		Section \ref{Sec:TrueSy2} (Fig. \ref{broadHa} green curve).}\\
		\tablefoottext{a}{Line flux in units of $10^{-17}\,\mathrm{erg/cm^{2}/s}$.}
		\tablefoottext{b}{Full Width at Half Maximum in units of $\mathrm{km.s^{-1}}$.}
		\tablefoottext{$\ast$}{The FWHM of the broad component is the fitted value when a weak 
		broad line is present, otherwise it is fixed at the predicted value (column 7 of the table).}
		}
	\end{table*}
		
	We can then compare the predicted and measured parameters for H$\alpha$. We can see 
	Fig. \ref{ObsvsMeasF} that there is a large difference 
	between the predicted and measured (or upper limit) fluxes of the broad H$\alpha$ component. For a 
	consistent observed flux with the prediction, we expect a value of $\Delta$ (with $\Delta = f_{BroadH
	\alpha,predicted}-f_{BroadH\alpha,measured}$) around 0. However, we can see that the minimum $
	\Delta$, even if considering the errors, is about 1000 (with a mean value $\sim 2000$), so far away 
	from the expected value of 0. Thus, there is no chance that the observed flux was consistent with the 
	predicted value for the broad H$\alpha$ line. The observed, or upper limit, broad component flux 
	constitutes only 0.5 to 2.4\% of the predicted flux. Moreover, these sources have an observed 
	broad H$\alpha$ flux $\gtrsim1$ dex lower than the predicted value. While the scatter in 
	the \citet{Greene05} $L_{Broad\,H\alpha}$-$\lambda L_{5100}$ relation, due to the intrinsic scatter of 
	QSO properties, is only $\sim0.2$ dex; so the observed difference between the observed and 
	predicted fluxed is a real effect. We thus conclude that a broad H$\alpha$ line would have 
	been securely detected if present at the expected level. In conclusion these sources clearly lack 
	broad lines and seem to be True Sy2.

\section{Summary and conclusions}
\label{sec:cl}
	We have investigated the nature of a sample of X-ray selected AGN optically (mis)classified as SF 
	galaxies in the BPT diagnostic diagram (elusive AGN), and have discussed possible reasons for 
	the misclassification. For this purpose, we have looked at their optical and X-ray properties.
	
	We find that 59\% of the elusive AGN are very likely to be NLS1, showing a steep X-ray 
	spectrum with no intrinsic absorption consistent with a low hardness ratio, forbidden line widths 
	smaller than the Balmer line widths and a high Eddington ratio with low mass black hole. It thus 
	seems that these sources are misclassified in the optical as SF galaxies because of the 
	additional contribution to the Balmer line flux from the BLR and NLR; while the BPT diagram 
	classification assumes contributions only from the NLR.
	
	The other 41\% of the remaining elusive AGN have the some properties of type 2 AGN, they 
	have only narrow emission lines in their optical spectra, X-ray luminosity of an AGN but some of them 
	show no sign of absorption (or little absorption), it is that why we call them \emph{X-ray Sy2}. Three 
	possible explanations for their optical elusiveness have been investigated. Heavy obscuration by 
	gas and dust is not a likely origin of the misclassification. However, two other 
	possibilities could explain the optical elusiveness of these sources. Firstly, 60\% of them are 
	optically diluted by starlight (so are weak AGN compared to the host galaxy) with a possible star-
	formation contribution in some cases. Secondly, the \emph{X-ray Sy2} sources also have low 
	accretion rates and are thus optically underluminous due to an intrinsic weakness of the AGN. \\
	Moreover, among these \emph{X-ray Sy2}, we find a population of True Sy2, e.g. undiluted 
	AGN but with a lack of broad emission lines in their optical spectra. The absence of broad lines has 
	been checked by comparing the observed optical spectra with the predicted flux and width of 
	the H$\alpha$ broad component. According to some theoretical models, their very low accretion rates 
	cause the disappearance of the BLR.

\begin{acknowledgements}
	We would like to thank Andrew Blain and Gordon Stewart, together with the referee, Jonathan 
	Trump, for their useful comments.
	\\This work uses data from observations obtained with XMM-Newton, an ESA science mission with 
	instruments and contributions directly funded by ESA Member States and NASA. Data from the 
	SDSS is also utilised.Funding for SDSS-III has been provided by the Alfred P. Sloan Foundation, 
	the Participating Institutions, the National Science Foundation, and the U.S. Department of Energy 
	Office of Science. The SDSS-III web site is http://www.sdss3.org/. This publication also makes use 
	of data products from the Wide-field Infrared Survey Explorer, which is a joint project of the 
	University of California, Los Angeles, and the Jet Propulsion Laboratory/California Institute of 
	Technology, funded by the National Aeronautics and Space Administration.
\end{acknowledgements}


\bibliographystyle{aa}
\bibliography{ref1}

\end{document}